\documentclass[preprint,letterpaper]{aastex}

\usepackage{amsmath, amssymb, graphicx, color}
\usepackage{natbib}

\allowdisplaybreaks

\def\reduce{\rotatebox{20}{{--}} \hspace{-0.55em} }
\newcommand{\doubledot}[1]{\raisebox{2ex}{\scriptsize ..} \hspace{-0.6em}{#1}}
\newcommand{\tripledot}[1]{\raisebox{2ex}{\scriptsize ...} \hspace{-0.65em}{#1}}
\def\sss{\scriptscriptstyle}

\begin{document}

\title{The Role of the Equation of State in Binary Mergers }
\author{Sa{\v s}a Ratkovi{\' c}}
\email{ratkovic@grad.physics.sunysb.edu}
\affil{Department of Physics \& Astronomy, \\
        State University of New York at Stony Brook, \\
        Stony Brook, NY 11794-3800, USA}
\author{Madappa Prakash}
\email{prakash@helios.phy.phiou.edu}
\affil{Department of Physics \& Astronomy, \\
       Ohio University,\\
       Athens, OH 55701, USA}
\author{James M. Lattimer}
\email{lattimer@mail.astro.sunysb.edu}
\affil{Department of Physics \& Astronomy, \\
        State University of New York at Stony Brook, \\
        Stony Brook, NY 11794-3800, USA}
\shorttitle{Binary mergers}
\shortauthors{Ratkovi{\' c}, Prakash, and Lattimer}

\begin{abstract}
Binary mergers involving black holes and neutron stars have been
proposed as major sources of gravitational waves, r--process
nucleosynthesis, and gamma ray bursters. In addition, they represent
an important, and possibly unique, observable that could distinguish
between normal and self--bound neutron stars. These two families of
stars have distinctly different mass--radius relationships resulting
from their equations of state which can be revealed during their
mergers if stable mass transfer ensues.  We consider two cases of
gravitational-radiation induced binary mergers: (i) a black hole and a
normal neutron star, and (ii) a black hole and a self-bound strange
quark matter star.  We extend previous Newtonian analyses to
incorporate the pseudo-general relativistic Paczy\'nski-Wiita potential
or a potential correct to second--order post-Newtonian order in
Arnowitt--Deser--Misner coordinates. These potentials are employed to
study both the orbital evolution of the binary and the Roche lobe
geometry that determines when and if stable mass transfer between the
components is possible. The Roche lobe geometry with pseudo-general
relativistic or post-Newtonian potentials has not heretofore been
considered.  Our analysis shows that differences in the evolution of
normal neutron stars and strange quark matter stars are significant
and could be detected in gravity waves. Both the amplitude and
frequencies of the wave pattern are affected. In addition, details of
the equation of state for either normal neutron stars or strange quark
stars may be learned. A single merger could reveal one or two
individual points of the mass-radius relation, and observations of
several mergers could map a significant portion of this relation.

\end{abstract}





\section{Introduction}

Mergers of compact objects in binary systems, such as a pair of
neutron stars (NS-NS), a neutron star and a black hole (NS-BH), or two
black holes (BH-BH), are expected to be prominent sources of
gravitational radiation \citep{THORNE1}.  The gravitational-wave
signature of such systems is primarily determined by the chirp mass
$M_{chirp}=(M_1M_2)^{3/5}(M_1+M_2)^{-1/5}$, where $M_1$ and $M_2$ are
the masses of the coalescing objects. The radiation of gravitational
waves removes energy which causes the mutual orbits to decay. For
example, the binary pulsar PSR B1913+16 has a merger timescale of
about 250 million years, and the pulsar binary PSR J0737-3039 has a
merger timescale of about 85 million years \citep{LYNE04}, so there is
ample reason to expect that many such decaying compact binaries exist
in the Galaxy.  Besides emitting copious amounts of gravitational
radiation, binary mergers have been proposed as a source of the
r-process elements \citep{LSCH2} and the origin of the
shorter-duration gamma ray bursters \citep{EICHLER89}.


The expected rates of binary mergers have been estimated by either
utilizing the observational information from known binaries or else
through theoretical models of binary stellar evolution (population
synthesis).  The coalescence rate for the NS-NS case has been recently
revised upward in view of the newly discovered binary PSR J0737-3039
\citep{LYNE04}. Due to its low luminosity and relatively short
lifetime, the estimated NS-NS merger rate has been increased from
prior estimates by almost an order of magnitude \citep{BURGAY03,
KALOGERA03}.  The merger rates can be coupled with expected
characteristics of future gravity wave detectors to make predictions
of the observed merger rate.  The updated analysis predicts the
observed NS-NS merger rate to be $1.8\times 10^{-4}$ yr$^{-1}$
galaxy$^{-1}$ with an uncertainty of about a factor of 3 to 10.  For
the advanced LIGO detector, it is estimated that mergers up to
distances of 350 Mpc \citep{FINN2001} could be observed, with a total
detection rate of up to approximately $1$ day$^{-1}$ anticipated
\citep{KALOGERA03}.  In the NS-BH case the merger rate has been
estimated to be about $10^{-5}-10^{-4}$ yr$^{-1}$ galaxy$^{-1}$
\citep{BETHE98,PORTEGIES98}, or up to 6 times larger than the NS-NS
merger rate.  It seems likely that this rate might also be adjusted
upward by the discovery of PSR J0737-3039.


A neutron star in a binary merger with a larger mass companion will be
tidally disrupted.  If the tidal disruption occurs far enough outside
the innermost stable circular orbit (ISCO) of the binary, it is expected
that an accretion disc will form or that mass transfer to its
companion will occur \citep{CLARK1,Jaranowski92,PORTEGIES98b}.
 In either case, the gravity wave signal is expected to be
significantly different than if tidal disruption occurs after the star
penetrates the ISCO.  Tidal disruption will occur at larger
separations for larger neutron star radii, so some aspects of the
neutron star equation of state (EOS) could be determined in the case
in which tidal disruption occurs outside the ISCO.

Depending upon the equation of state (EOS) of dense matter and the
mass ratio of the binary, tidal disruption could lead to the less
massive star overflowing its Roche lobe and transferring its mass
through the inner, or first, Lagrange point to its more massive
companion.  If mass loss from the lighter star occurs quickly enough,
and if conservation of mass and orbital angular momentum can be
assumed, the two stars will then begin to spiral apart.  Although this
increases the Roche lobe volume, the radius of the neutron star also
increases in response to the mass loss.  Mass transfer will continue
in a stable fashion if the lighter star can expand sufficiently fast
such that it is able to continuously fill its Roche surface.  We refer
to mass transfer under such conditions as {\em stable mass transfer}.
Because the orbital separation now increases, the gravity wave
amplitude will decrease.  The signature of stable mass transfer in
gravity waves should therefore be strikingly different than for an
amorphous tidal disruption or a direct plunge.  As we show below,
important information about the radius of a neutron star and the
underlying equation of state could be contained in the gravity wave
signal. The longer timescale produced by stable mass transfer might also
extend the duration of an associated gamma ray burst \citep{PORTEGIES98b}.
In the absence of mass transfer, this timescale might be of
order milliseconds, the orbital period of the binary near the last
stable orbit.  In contrast, stable mass transfer extends over a period
of at least several tenths of a second.

Another effect of stable mass transfer would be to modify
the amount of material potentially ejected from the system.  Matter
from a tidally disrupted neutron star, which could be accelerated to
escape velocities from the binary \citep{LSCH2}, undergoes
decompression which results in heavy nuclei and an intense neutron
flux leading to the copious production of r-process elements
\citep{LATTIMER77,MEYER89}. 
In the case in
which stable mass transfer occurs, sudden disruption of the neutron
star near the last stable orbit is avoided, but mass could be
transferred unstably at later times and larger separations when the
neutron star approaches its minimum stable mass \citep{COLPI93}.


One of the most significant aspects of the equation of state that
stable mass transfer could reveal is whether or not the neutron star
is actually a strange quark matter star.  It has been suggested that
if strange quark matter is the ultimate ground state of matter (i.e.,
has a lower energy at zero pressure than iron) compression of neutron
star matter to sufficiently high density triggers a phase transition
which converts virtually the entire neutron star to strange quark
matter \citep{MADSEN99}.  Such a star is self-bound as opposed to
being gravitationaly bound as is the case of a normal neutron star.

It has so far proved very difficult to find venues from astrophysical
observations that could unambiguously distinguish strange quark stars
from normal neutron stars.  This is because self-bound stars have
similar radii, moments of inertia, and neutrino emissivities and
opacities to that of moderate mass normal neutron stars.  Therefore,
it may be unlikely that photon or neutrino observations, or radio
binary pulsar timing measurements, will be able to differentiate these
cases, especially if strange quark stars have a small hadronic crust,
supported perhaps by electrostatic forces.  In that case, the
effective temperatures and radii of solar-mass-sized strange quark
stars and normal neutron stars would tend to be similar.  Even during
the proto-neutron star stage, which is observable through neutrino
emissions \citep{Burrows86}, these two types of stellar configurations
likely yield similar neutrino signals until such late times that the
low luminosities prevent an unambiguous discrimination
\citep{Prakash01}.

However, major differences in the evolutions of normal neutron stars
and strange quark matter stars emerge during the final stages of
binary mergers if stable mass transfer occurs. These differences would
be prominent in both the amplitudes and frequencies of gravitational
wave emisions.


The Newtonian equations for orbital motion become progressively
inadequate as compact objects spiral inward.  For example, the
existence and location of the innermost stable circular orbit (ISCO)
are not predicted by Newtonian gravitation.  Furthermore, in
calculations performed to date, the gravitational equipotential
surfacws and the size of the Roche lobe
have been computed using a Newtonian background
\citep{KOPAL1,PACZYNSKI1,EGGLETON1}). Although a solution in full GR
is not yet possible, systematic post-Newtonian
corrections have been developed to explore regions close to compact
objects (see, for example, \citep{BLANCHET3}).  To extend the
Roche lobe overflow model for mass transfer, we will employ recently
calculated corrections to the Newtonian results by using the second
order post--Newtonian approximation in Arnowitt--Deser--Misner
coordinates \citep{ROCHE2PN}.


We adopt a semi-analytic approach to describe the final stages of
NS-NS or BH-NS mergers, extending earlier treatments
\citep{Prakash03}.  While the evolution of such mergers has been
previously calculated with numerical techniques involving three-dimensional
Newtonian hydrodynamics \citep{KLUZNIAK1}, only a few
simulations have implemented GR corrections to the orbital evolution
or have considered the case in which the inspiralling stars have
unequal masses \citep{DAVIES2005,BEJGER05,SHIBATA05a,SHIBATA05b}.  In
the final phase of a merger, Newtonian orbital mechanics cannot be
expected to hold and general-relativistic (GR) corrections become
important. In the few cases in which unequal masses were considered,
the mass ratios were severely restricted because of numerical
difficulties \citep{Shibata03}.


This work is organized as follows.  In \S \ref{sec:General}, we
provide the general considerations upon which this work is based on
and discuss the mechanism of Roche lobe overflow. The three main
ingredients: (1) the evolution of orbital angular momentum, (2) the
evolution of the Roch lobe radius, and (3) the EOS parameter that
governs these evolutions, are then presented. The differential
equations that govern the evolution of the separation distance $a$ and
the mass ratio $q=M_1/M_2$ are derived in this section along with the
condition for stable mass transfer.  In \S 3, the EOS's of normal
neutron stars and self-bound quark stars used in this work are
summarized.  \S 4 provides the evolution equations for mergers using
the second order post--Newtonian potential.  We summarize here the
recently calculated results for Roche lobes in the second order
post--Newtonian approximation \citep{ROCHE2PN}.  In \S 5, we
introduce, in addition to the Newtonian and 2PN analyses, two
pseudo--general relativistic potentials for comparisons of binary
evolution simulations. We first introduce the potential due to
\citet{PACZYNSKI2}, and then we modify it to construct a hybrid potential that
correctly incorporates some post--Newtonian features.  We fit its
parameters to match predictions for the position of the innermost
stable circular orbit (ISCO) from \citet{BLANCHET2003}, who evaluated
the effective gravitational potential up to order $(v^2/c^2)^3$ in
post-Newtonian order.  We extend the Newtonian analysis of Roche lobes
to include the Paczy\' nski--Wiita \citep{PACZYNSKI2} 
and hybrid potentials and provide
simple fitting results.  We consider the corresponding changes to the
orbital dynamics that affects the gravitational radiation reaction and
the location of the ISCO.  Regions of stable mass transfer in the
$M_1-M_2$ plane are identified in \S 6 for all potentials
considered.  In \S \ref{sec:evolution}, we present results of merger
simulations for both a normal star and a self-bound star orbiting
around another more massive object (BH or NS).  A variety of equations
of states with varying stiffness at supranuclear density for both
cases are considered.  Our results pertaining to
gravity wave signals are presented in \S 8.  Conclusions are 
contained in \S \ref{sec:conclusion}.


\section{General considerations}
\label{sec:General}

Our objective is to explore the role of the EOS
as it affects compact binary mergers (see also
\citet{Lee01,FABER02,Prakash03,PRAKASH04,Shibata03,DAVIES2005,BEJGER05,
SHIBATA05a,SHIBATA05b}). In general, a gravitational merger is the
result of a system which contains two objects in a mutual orbit
decaying via gravitational radiation reaction.  When the separation
becomes small enough, the less massive object can exceed its Roche
lobe and begin to transfer mass to the more massive companion.  Typically,
this can occur if the mass ratio $q=M_1/M_2$ of the two objects ($M_1$
is the neutron star mass) is somewhat less than unity.  As mass is
transferred, the radius of the neutron star readjusts to its new mass
on hydrodynamic timescales, which are rapid compared to timescales of
orbital evolution.  Normally, the radius of a neutron star expands as
its mass is lost, but there are exceptions.  For example, a self-bound
quark star which is not near its maximum mass has a radius which {\it
increases} as the cube root of its mass.  But even if this is the
case, mass transfer can continue because the Roche lobe radius tends
to shrink as $q$ decreases.  Even though conservation of angular
momentum during mass transfer will reverse the binary's inspiral,
thereby causing an increase in the Roche lobe radius, it is still
possible the radius increase can keep pace.  If this occurs, the mass transfer
is stable.  During stable mass transfer, the stellar radius and its
Roche lobe remain coincident and mass transfer can continue until the
star's mass becomes very small.  For normal neutron stars, stable mass
transfer terminates when the star's mass approaches the minimum
neutron star mass of approximately 0.1 M$_\odot$, but for strange
quark matter stars, it continues indefinitely.

In our treatment of the star's orbit, the stars are considered to be
point-like masses and the angular momentum of the system is assumed
not to be transfered into the spins of the two compact objects. The
orbital angular momentum changes only through the emission of gravity
waves.  We also assume that the total mass of the system is conserved
so that mass transfer only exchanges mass between the two bodies and
not, for example, into an accretion disc or into material ejected from
the system.  We do not expect that these
assumptions invalidate the qualitative value of our analysis, but
adopting a more general treatment might alter some quantitative
features such as the gravity wave signal amplitude.

There are four main ingredients in our analysis. First, we assume that
mass is conserved during mass transfer.  Thus, using $M=M_1+M_2$, 
\begin{eqnarray}
\frac{\dot{M}_1}{M_1}
&=&
\frac{1}{1+q}\, \frac{\dot{q}}{q}\, .
\end{eqnarray}
In the case of fixed total mass $M$, the change of
the total angular momentum $J$ depends only on the change of the mass
ratio $q$ and the orbital separation $a$.  Second, we assume that the
change of $J$ is balanced by the loss of angular momentum $J_{GW}$
that the outgoing gravitational waves carry:
\begin{eqnarray}
\frac{dJ}{dt}
&=&
\frac{\partial J}{\partial a} \dot{a}
+
\frac{\partial J}{\partial q} \dot{q}
= - \dot{J}_{GW}\, .
\label{EQ:BASICJ}
\end{eqnarray}
The second ingredient is related to the equipotential surface that
regulates mass transfer. The effective sizes of the Roche lobes also
depend, for a fixed total mass, only on $q$ and $a$. Hence, the time
dependence of the Roche lobe radius $r_{\sss Roche}$ can be written as
\begin{eqnarray}
\dot{r}_{\sss Roche} 
&=&
\frac{\partial r_{\sss Roche}}{\partial a} \dot{a}
+
\frac{\partial r_{\sss Roche}}{\partial q} \dot{q}\, .
\label{EQ:BASICROCHE}
\end{eqnarray}

The last ingredient incorporates the dependence of binary mergers on
the EOS of dense matter.  We define a parameter $\alpha$ describing how the
radius of the star changes as mass is stripped from it:
\begin{eqnarray}
\alpha(M_1) &\equiv&
\frac{d \ln R_1}{d \ln M_1} \,.
\label{eq:alphadef}
\label{EQ:BASICEOS}
\end{eqnarray}

During stable mass transfer the star's radius remains equal
to the Roche radius, that is, $R_1 = r_{\sss Roche}$. We can combine
this equality with equations (\ref{EQ:BASICROCHE}) and
(\ref{EQ:BASICEOS}) to establish a connection between the
derivatives $\dot{a}$ and $\dot{q}$
\begin{eqnarray}
\frac{\partial \ln r_{\sss Roche}}{\partial \ln a}
\frac{\dot{a}}{a}
+
\frac{\partial \ln r_{\sss Roche}}{\partial \ln q}
\frac{\dot{q}}{q}
&=&
\frac{\alpha(M_1)}{1+q}\, \frac{\dot{q}}{q}\, ,
\end{eqnarray}
which can be expressed as
\begin{eqnarray}\label{EQ:QDOT}
\frac{\dot{q}}{q} - \Upsilon(a,q)\, \frac{\dot{a}}{a} 
&=&
0\, ,
\end{eqnarray}
where
\begin{eqnarray}
\Upsilon(a,q) 
&\equiv& 
\frac{\displaystyle
\frac{\partial \ln r_{\sss Roche}}{\partial \ln a}
}
{\displaystyle
\frac{\alpha(M_1)}{1+q}
-
\frac{\partial \ln r_{\sss Roche}}{\partial \ln q}
}\, .
\label{EQ:UPSILON}
\label{eq:chi1}
\end{eqnarray}
We supress the dependence of variables on the total mass $M$.
Generally, one has \\ $\partial\ln r_{Roche}/\partial\ln a\approx1$ and
$\partial\ln r_{Roche}/\partial\ln q\approx 1/3$.  Coupled with the
fact that, in general, $\alpha\le1/3$ (see \S 6), one has  
$\Upsilon(q, a)<0$. 

These results can be combined with equation (\ref{EQ:BASICJ}) in order
to find the set of coupled differential equations that govern the
evolution of $a$ and $q$.  The separation distance evolves in time
according to
\begin{eqnarray}
\dot{a} 
&=&
- \, \frac{\dot{J}_{GW}(a, q)}
{ \displaystyle
\frac{\partial J(a, q)}{\partial a} +
\Upsilon(a,q) \, \frac{q}{a} \, \frac{\partial J(a, q)}{\partial q} } \, ,
\label{EQ:DIFFA}
\end{eqnarray}
and simultaneously the mass ratio changes according to
\begin{eqnarray}
\dot{q} 
&=&
- \, \frac{\dot{J}_{GW}(a, q)\, \Upsilon(a, q)}
{ \displaystyle
\frac{a}{q} \, \frac{\partial J(a, q)}{\partial a}  +
\Upsilon(a,q)\, \frac{\partial J(a, q)}{\partial q} }\, .
\label{EQ:DIFFQ}
\end{eqnarray}
In stable mass transfer, $M_1$ decreases so $q$ does also.
Furthermore, it is generally true that \\ $\partial J/\partial a>0$ and
$\partial J/\partial q>0$.  Thus, the condition for stable mass
transfer is simply
\begin{equation}
\Upsilon(a,q)>
-
\frac{\partial\ln J(a,q)/\partial\ln a}
{\partial\ln  J(a,q)/\partial\ln q}\,.
\label{EQ:STABLE}
\end{equation} 
During stable mass transfer,  $\dot a>0$ and, hence, the
binary spirals apart.  However, if this condition is subsequently
violated, mass transfer either ceases or becomes unstable.
If mass transfer terminates, emission of
gravitational radiation leads to a contraction of the orbit, whereupon
mass transfer may be able to begin again.  In either case, this limit
spells a quick end to the star.

Equations (\ref{EQ:DIFFA}) and (\ref{EQ:DIFFQ}) determine the time
evolution of $a$ and $q$. From these quantities, all other relevant
quantities can be determined during mass transfer.  The details will
depend upon the EOS and upon the gravitational potential assumed.  In
the following sections, we explore several gravitational potentials and
corresponding models for the Roche radius to study the behavior of
binary systems that are built of stars with different equations of
state.

We anticipate that tidal distortions, which give rise to the Roche
geometry, do not significantly alter the orbital evolution, because
the associated gravitational radiation corrections are small (see
\citet{CHAU1} and \citet{CLARK2}).  We also restrict ourselves to the
study of conservative mass transfer (in which angular momentum is
conserved).


\section{The equation of state}
\label{sec:EOS}

Several proposals concerning the physical state and the internal
constitution of matter at supra-nuclear densities have been put forth
(see~\cite{LP01,Alford01} for recent accounts).  The 
many exciting possibilities for the composition of compact stars
include (1) strangeness-bearing matter in the form of hyperons,
kaons, or quarks, (2) Bose (pion or kaon) condensed matter, and (3)
so-called self-bound strange quark matter (SQM).  Fermions, whether
they are in the form of baryons or deconfined quarks, are expected to
additionally exhibit superfluidity and/or superconductivity. 
The possibility of such exotic phases brings attendant changes to the
predictions of maximum masses and radii, since the presence of
multiple components or new phases of matter generally lessens the
pressure for a given energy density. For the discussion at hand, we
will use the term normal star to refer to a star with a surface of
normal matter in which the pressure vanishes at vanishing baryon
density. The interior of the star, however, may contain any or a
combination of the above exotica.  

A self-bound star is exemplified by Witten's SQM star
\citep{WITTEN84}; see \citet{Alcock} for a review.  Such a star has a
bare quark matter surface in which the pressure vanishes at a finite
but supra-nuclear baryon density. In the context of the MIT bag model
with first order corrections due to gluon exhange, the baryon density
at which pressure vanishes is
\begin{equation}
n_b(P=0) =
(4B/3\pi^{2/3})^{3/4}(1-2\alpha_c/\pi)^{1/4} \,,
\end{equation}
where $B$ is the bag
constant and $\alpha_c=g_c^2/(4\pi)$ is the quark-gluon coupling
constant. This density is not significantly affected by
the finite strange quark mass~\citep{PBP90} or by the pairing
phenomenon in quark matter ~\citep{JMad}.

\begin{figure}[!hbt]
\begin{center}
\epsscale{1.0}
\plotone{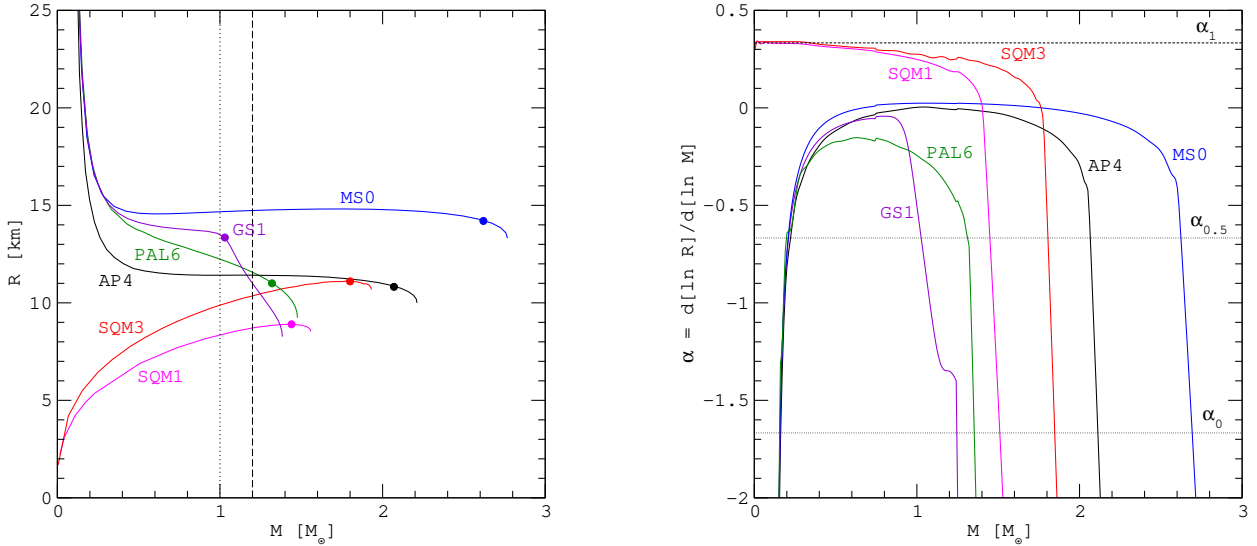}
\end{center}
\caption
[Radius and its logarithmic derivative as functions of stellar mass
for prototype EOS's.]
{\label{fig:rmalpha}Radius versus mass (left panel) and its
logarithmic derivative (right panel) for prototype EOS's. The EOS
symbols are as in \citet{LP01}.  The vertical lines in the left
panel and the horizontal lines in the right panel are discussed in \S
6. In the left panel, the maximum masses for which stable mass
transfer is allowed in the case $q=0.5$ are denoted by filled
circles for each EOS. }
\end{figure}

Prototypes of the radius versus mass for broadly differing EOS's
selected from \citet{LP01} are shown in Figure
\ref{fig:rmalpha}. Table \ref{TBL:EOS} summarizes the nomenclature
used in this figure as well as the masses and radii of the maximum
configurations of stars constructed using these EOS's.  Quantitative
variations from these generic behaviors can be caused by uncertainties
in the underlying strong interaction models (see the compendium of
results in Figure 2 of \cite{LP01}). Qualitative differences in the
outcomes of mergers with a black hole emerge, however, because of the
gross differences in the mass-radius diagram.

\begin{deluxetable}{l l l l l}
\tablecaption{EQUATIONS OF STATE}
\tablewidth{0pt}
\tablehead{ \colhead{Symbol} & \colhead{Approach} &
\colhead{Composition} & \colhead{$M_{max} [M_\odot$]} &
\colhead{$R(M_{max})$ [km]} } \startdata AP4\tablenotemark{a} &
Variational & np & 2.21 & 10.0 \\ MS0\tablenotemark{b} & Field
Theoretical & np & 2.77 & 13.32 \\ PAL6\tablenotemark{c} & Schematic
Potential & np & 1.48 & 9.23 \\ GS1\tablenotemark{d} & Field
Theoretical & npK & 1.38 & 8.27 \\ SQM1 (SQM3)\tablenotemark{e} &
Quark Matter & Q $(u,d,s)$ & 1.56 (1.93) & 8.54 (10.69) \enddata
\tablecomments{Approach refers to the underlying theoretical
technique.  Composition refers to strongly interacting components
(n=neutron, p=proton, H=hyperon, K=kaon, Q=quark); all models include
leptonic contributions. The last two columns contain the corresponding
masses and radii of the maximum mass configurations.  }
\tablenotetext{a}{Akmal \& Pandharipande (1997)}
\tablenotetext{b}{M\"uller \& Serot (1996)} \tablenotetext{c}{Prakash,
Ainsworth \& Lattimer (1988)} \tablenotetext{d}{Glendenning \&
Schaffner-Bielich (1999)} \tablenotetext{e}{Prakash, Cooke \& Lattimer
(1995)}
\label{TBL:EOS}
\end{deluxetable}

A normal star and a self-bound star represent two quite different
possibilities (see the right panel in Figure \ref{fig:rmalpha}) for
the quantity
\begin{eqnarray}
\alpha \equiv  \frac {d\ln R}{d\ln M} 
\left\{ \begin{array}{ll} 
\leq 0 & \mbox{{\rm for~a~normal~neutron~star~(NS)}} \\
\ge 0 & \mbox{{\rm for~a~self-bound~SQM~star}} 
\end{array} 
\right. \,, 
\label{lderiv}
\end{eqnarray}
where $M$ and $R$ are the star's mass and radius, respectively. For
small to moderate mass self-bound stars, $R \propto M^{1/3}$ so that
$\alpha \cong 1/3$; only for configurations approaching the maximum mass
does $\alpha$ turn negative.

Our objective is to explore the astrophysical consequences of these
distinctive behaviors in $R$ versus $M$ as they affect mergers with a
black hole.  Note that $\alpha$ is intimately connected with the dense
matter equation of state (EOS), since there exists a one-to-one
correspondence between $R(M)$ and $P(n_B)$, where $P$ is the pressure
and $n_B$ is the baryon density.  Gravitational mergers in which a
compact star loses its mass (either to a companion star or to an
accretion disk) during evolution is one of the rare examples in which
the $R$ versus $M$ (or equivalently, $P$ versus $n_B$) relationship of
the same star is sampled.  Although we focus here on the coalescence
of a compact star with a BH, the theoretical formalism and our
principal findings apply also to mergers in which both objects are
compact stars.

\section{Mergers with the second order post--Newtonian potential}
\label{sec:2PN}

Under moderately strong gravitational fields, the post-Newtonian
approximation has been widely used to study systems in which general
relativistic corrections are considered to be important.
\citet{ROCHE2PN} have recently calculated corrections to the Newtonian
results for the Roche lobe geometry by using the second order
post--Newtonian approximation in the Arnowitt--Deser--Misner
coordinates.  Starting from the N-body Lagrangian derived by
\citet{DAMOUR85}, \citet{ROCHE2PN} developed the Lagrangian for a test
particle in the vicinity of two massive compact objects.  This
calculation required the use of an approximate expression for the
transverse--traceless term $U_{TT}$ of the Lagrangian, since an exact
result is not available. The approximate form for $U_{TT}$ that is
valid in the vicinity of the less massive star was established.  Next,
the 2PN effective potential in the co--rotating frame was found and in
a similar fashion as for the Newtonian case, the resulting values of
the effective Roche lobe radius as functions of $q$ and $a$ could be
determined.  Technically, the post-Newtonian Roche radii can be
adequately parametrized by the quantities $q$, $a$, and the relativity
parameter $z=2M/a$, the dimensionless ratio of the effective event
horizon of the total mass and the semi-major axis.  In the case in
which the total mass $M$ is assumed conserved, $z$ varies inversely
with $a$.  The results can thus be presented in a form that closely
resembles Newtonian expressions that depend only upon $a$ and $q$.

\subsection{Analytic fits to Roche lobe radii to 2 PN}
\label{sec:2PNRL}

In Newtonian gravity, the potential surfaces are proportional to $a$
and otherwise a function of only the mass ratio $q$, and are to a
sufficient accuracy often described by the Eggleton approximation
\citep{EGGLETON1}:
\begin{eqnarray}\label{eq:roche2pn}
  \frac{r_{\sss Roche}}{a} & = & Q(q)\equiv  
\frac{\alpha_{\sss Q}\, q^{2/3} } {\beta_{\sss Q}\, q^{2/3} 
+ \ln(1 + q^{1/3}  )}\, , 
\label{eq:q}
\end{eqnarray}
with $\alpha_Q=0.49$ and $\beta_Q=0.57$ as parameters.  Here, $r_{Roche}$
is the radius of a sphere of the same volume $V_{Roche}=4\pi
r_{Roche}^3/3$ of the equipotential surface which extends through the
first, or inner, Lagrangian point.  We will assume that Roche lobe
overflow occurs when the radius of the star $R_1$ exceeds the
effective Roche radius $r_{Roche}$, even though the true equipotential
surface is not spherical.  We will also assume that tidal deformations of the
stars do not affect the Roche picture of mass transfer significantly.

\begin{figure}[ht!]
\begin{center}
\includegraphics[width=0.9\textwidth]{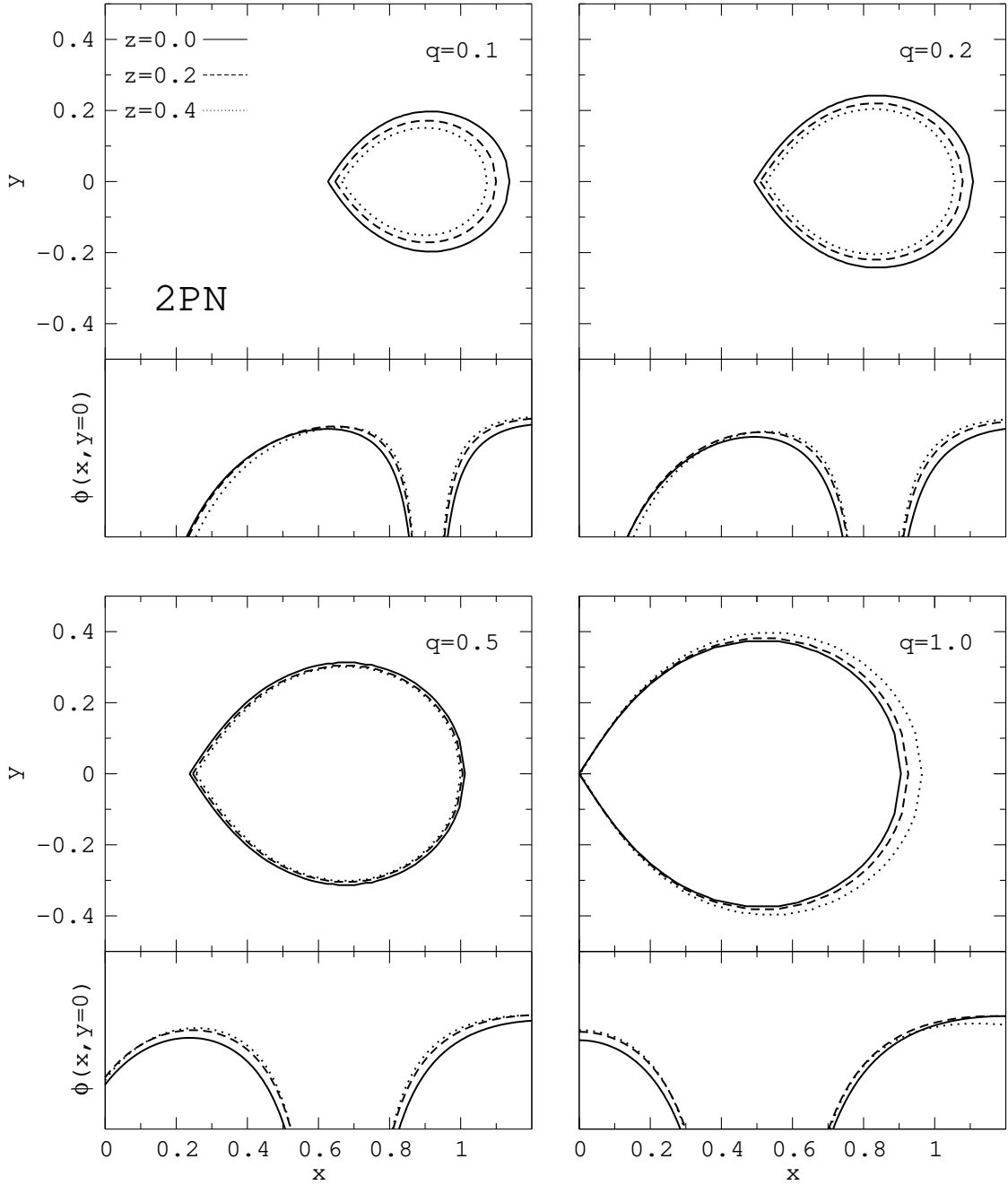}
\caption
[Roche lobes and effective potentials at the 2PN level.]
{Roche lobes and the corresponding potentials for $y=0$ for the second
  order post-Newtonian (2PN) potential. Coordinates $x$ and $y$ are scaled
  by the stellar separation $a$ and are shown for $q = 0.1$, $0.2$,
  $0.5$, and $1.0$, respectively. Results are shown for values of
  $z=0$, $z = 0.2$, and $z=0.4$, respectively.
\label{FIG:EQUIP2PN}}
\end{center}
\end{figure}

\begin{figure}[ht!]
\begin{center}
\includegraphics[width=0.6\textwidth]{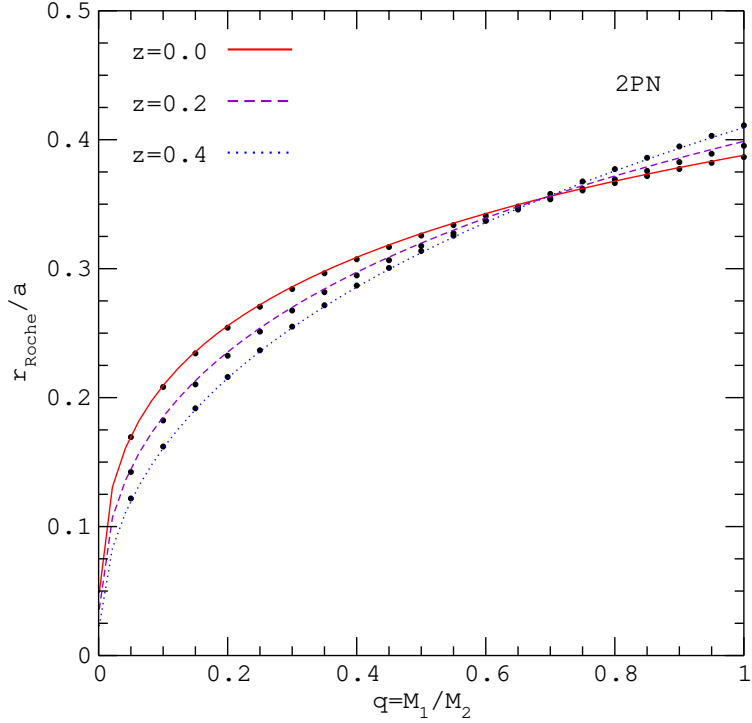}
\caption
[The effective Roche lobe radii at the 2PN level.]
{Effective Roche lobe radii $r_{\sss Roche}$ for the 2PN potential
scaled by the stellar separation $a$ versus mass ratio $q$
for $z=0$, $z = 0.2$, and $z=0.4$, respectively.
\label{FIG:ROCHE2PN}}
\end{center}
\end{figure}

To extend the Newtonian treatment to the post-Newtonian case,
an adequate approximation 
was constructed by \citet{ROCHE2PN}  
by utilizing an additional
relativity parameter $z=2M/a$ corresponding to the ratio of the
effective event horizon of the total mass and the orbital separation:
\begin{eqnarray}
\frac{r_{Roche}}{a}&=&Q(q)C(q,z)\,,
\label{eq:reqn}
\label{eq:approx}
\end{eqnarray}
where
\begin{eqnarray}
 C(q, z) & = & 1 + z\,(\alpha_{\sss C}\, q^{1/5} - \beta_{\sss C})
\label{eq:c}
\end{eqnarray}
with $\alpha_{\sss C} = 1.951$ and $ \beta_{\sss C} = 1.812$.  We
emphasize that the $z$ dependence is equivalent to an $a$ dependence
when $M$ is constant.  The Roche equipotential surfaces (which all
scale with the orbital separation $a$) in the equatorial plane along a
line through the stars' centers of mass are displayed in Figure
\ref{FIG:EQUIP2PN} for a variety of $q$ and $z$ values.  The results
for the Roche lobe radii, and the fits we adopted from
\citet{ROCHE2PN}, are shown in Figure \ref{FIG:ROCHE2PN}.  Note the
common point $q=(\beta_C/\alpha_C)^5\simeq0.69$ in these results.

\subsection{Evolution equations}
\label{sec:2PNEVOL}
In order to model the orbital evolution of the merging stars, it is
necessary to develop an  appropriate expression for the orbital
angular momentum $J$.  Although the angular momentum in the ADM
approximation has been derived up to the 3PN order
\citep{BLANCHET2001,BLANCHET2003}), for consistency we use results up
to the 2PN level. For quasi-circular motion, the angular momentum is
\begin{eqnarray}
J 
&=&
\sqrt {a}{M}^{3/2}
\frac{q}{\left( 1+q \right) ^{2}} 
\left\{ 
1
+2\,{\frac {M}{a}}
-\frac{1}{16}\, 
\frac{48 + 79\,q + 104\,{q}^{2} + 79\,{q}^{3} + 48\,{q}^{4}}{\left( 1+q \right) ^{4}}
{\left(\frac{M}{a}\right)}^{2}
\right\} \, ,
\label{EQ:2PNJ}
\end{eqnarray}
where $M=m_1+m_2$. The post-Newtonian correction is given  by the
second and third terms.  Under the assumption of conserved total
mass, the angular momentum is only a function of $a$ and $q$ even in
this case. The angular momentum loss
due to gravitational wave emission (see equation (4.9) and
related discussion in \citet{BLANCHET2003}) is, in the same approximation,
\begin{eqnarray}
\frac{d J}{d t}
&=&
\frac{\partial J}{\partial a} \dot{a} +
\frac{\partial J}{\partial q} \dot{q} 
=
-\dot{J}_{GW} = - \frac{32}{5} \, 
\frac{q^2}{{(1+q)}^4} \,
\frac{M^{3/2}}{a^{1/2}} \,
{\left(\frac{M}{a}\right)}^3 \, 
+
\mathcal{O}\left[\left(\frac{M}{a}\right)^4\right]
\, ,
\label{EQ:2PNBALANCE}
\end{eqnarray}
where the $\mathcal{O}\left(\left(\frac{M}{a}\right)^4\right)$ terms
will be neglected.  The partial derivatives with respect to $a$ and
$q$ in equation (\ref{EQ:2PNJ}) are
\begin{eqnarray}
\frac{\partial J}{\partial a}
&=&
\frac{1}{2}\,\frac {{M}^{3/2}}{ {a}^{1/2}}
\,\frac{q}{{(1+q)}^2}
\Bigg\{
1
-2\frac {M}{a}
+ \frac{1}{16} \, {\left({\frac{M}{a}}\right)}^2
\,
\frac{144 + 237 q + 312 q^2 + 237 q^3 + 144 q^4}{{(1+q)}^4} 
\Bigg\}\, ,
\end{eqnarray}
and
\begin{eqnarray}
\frac{\partial J}{\partial q}
&=&
a^{1/2}\, M^{3/2}\, \frac{1-q}{{(1+q)}^3}
\Bigg\{
1
+2 \, \frac{M}{a}
+ \frac{1}{8} \, {\left( \frac{M}{a} \right)}^2 \,
\frac{-3 + 101q + 145q^2 +101q^3 - 3q^4}{{(1+q)}^4}
\Bigg\}\, .
\end{eqnarray}

We are now equipped to investigate for what choices of masses
$M_1$ and $M_2$ (i) stable mass transfer will occur, and (ii) whether
mass transfer will begin before the ISCO is reached.  Before
performing this analysis, however, we investigate two alternative
approaches to extending the Newtonian analysis to the relativistic
regime.  Neither of these alternate models have been fully
explored with the semi-analytic approach we will utilize for the 2PN
methodology described above, so we do this in the next section.


\section{Mergers with pseudo-GR potentials}
\label{sec:pseudopotentials}

As a simple alternative analysis that can be easily tailored to fit
some aspects of GR, the Paczy\' nski--Wiita potential
\citep{PACZYNSKI2} was used by \citet{KLUZNIAK1}. In our work, to
mimic the GR effects, we also use this family of pseudo-potentials
which replace the Newtonian potential $\phi_N(r)=-M/r$.  We endeavor
that pseudo-GR potentials accurately predict the existence of the ISCO
for the general case. The Paczy\' nski--Wiita potential correctly
predicts its location for the case of two arbitrary point masses. We
introduce an improved form of this potential, the hybrid
pseudo-general relativistic (henceforth pseudo-GR) potential, in order
to fit this potential more closely to post--Newtonian results in the
case of two noninfinitesimal masses.

\subsection{Paczy\' nski--Wiita potential}

Pseudo-potentials have proved useful for incorporating general
relativistic effects in the study of accretion disks surrounding a
massive central body. For test particles near a highly relativistic
object, \citet{PACZYNSKI2} introduced a pseudo-GR  potential
which we will refer to as the PW potential:
\begin{eqnarray}\label{eq:paczynskypot}
  \phi_{\scriptscriptstyle PW} (r) &=& -~\frac{M_2}{r-r_{\sss G2}}\, ,
\end{eqnarray}
where $M_2$ is the mass of the central compact object, $r$ is the
distance to the center of this object, and $r_{\sss G2}=2M_2$ is its
Schwarzschild radius. Applied to a test particle, this potential both
mimics the existence of an event horizon at $r=r_{\sss G2}$ and
correctly predicts that the ISCO is at $r = 3r_{\sss G2}$.  For other
examples of pseudo--potentials, see \citet{Lee01}.

\subsection{Hybrid pseudo-GR potential}
\label{SEC:ZETA}
The location of the ISCO for the PW potential agrees with the GR
result only for an orbiting particle with negligible mass.  However,
with slight modification, the point of gravitational instability of
the Paczy\' nski--Wiita potential can be positioned at the desired
radius. For a binary consisting of two nearly equal masses,
post-Newtonian results \citep{BLANCHET2003} valid up to 3PN order
predict that the ISCO moves inward compared to the point--particle result
($3r_{\sss G}$).  We can simulate these results by minimally modifying
the PW potential. Explicitly, we adopt the hybrid potential
\begin{eqnarray}\label{eq:hybridpot}
  \phi_{\scriptscriptstyle H} (r) &=& -~\frac{M}{r-\zeta(q)r_{\sss G}}\, ,
\label{eq:hybrid}
\end{eqnarray}
where $M=M_1+M_2$ is the total mass, $r_{\sss G}=2M$ is the sum of the
horizon radii of the two stars, and the function $\zeta(q)\in(0, 1)$
contains the necessary corrections to account for the position of the
ISCO as a function of the mass ratio $q=M_1/M_2$ of the two
components. We take $M_1$ to be the less massive star, so that $q<1$.
In the case of a test particle, the function $\zeta(0)=1$, whereas for
finite $M_1$, the correction function $\zeta(0\le q \le 1)<1$.  With
this hybrid potential, the orbital separation corresponding to the
ISCO becomes
\begin{eqnarray}
r_{ISCO}=3\zeta(q)r_{\sss G}.
\end{eqnarray}

We require that the hybrid potential predict the ISCO as closely as
possible to post-Newtonian results up to third order.  As shown by
\citet{BLANCHET2003}, the ISCO radius at which the gravitational
instability occurs for a binary system with a mass ratio $q$ can be
found in the 3PN approximation by solving the cubic equation resulting
from extremizing the gravitational potential,
\begin{eqnarray}\label{EQ:ZETACUBIC}
  1 + \alpha (\nu)x + \beta (\nu)x^2 + \gamma (\nu)x^3 &=& 0 \, 
\end{eqnarray}
with coefficients 
\begin{eqnarray}
\alpha(\nu) & = & -9 + \nu\, ,  \quad
\beta(\nu) = \frac{117}{4} + \frac{43}{8}\nu + \nu^2 \, , 
\quad {\rm and} \nonumber \\
\gamma(\nu) 
&=& 
-61 + \left(\frac{135403}{1680}-\frac{325}{64}\pi^2 -
\frac{99}{3}\lambda \right)\nu - \frac{31}{8}\nu^2 + \nu^3 \, ,
\end{eqnarray}
where $\nu=q/(1+q)^2$.  The quantity $\lambda$ arises from
considerations of dimensional regularization, and following
\citet{BLANCHET2003}, we set $\lambda=0.64$.  The ISCO radius is then
$r_{ISCO}=M/x_0(\nu)$, where $x_0(\nu)$ denotes the solution of the
cubic equation \ref{EQ:ZETACUBIC}.  We find that the 3PN corrections can
be fitted by a function $\zeta_{\scriptscriptstyle FIT}$ that
is a third order polynomial in $q$
\begin{eqnarray}\label{eq:zeta_fit}
  \zeta_{\scriptscriptstyle FIT}(q) &=& 0.697 + 0.214 {(1.07-q)}^3\,.  
\end{eqnarray}

\begin{figure}[!ht]
\begin{center}
\includegraphics[width=0.6\textwidth]{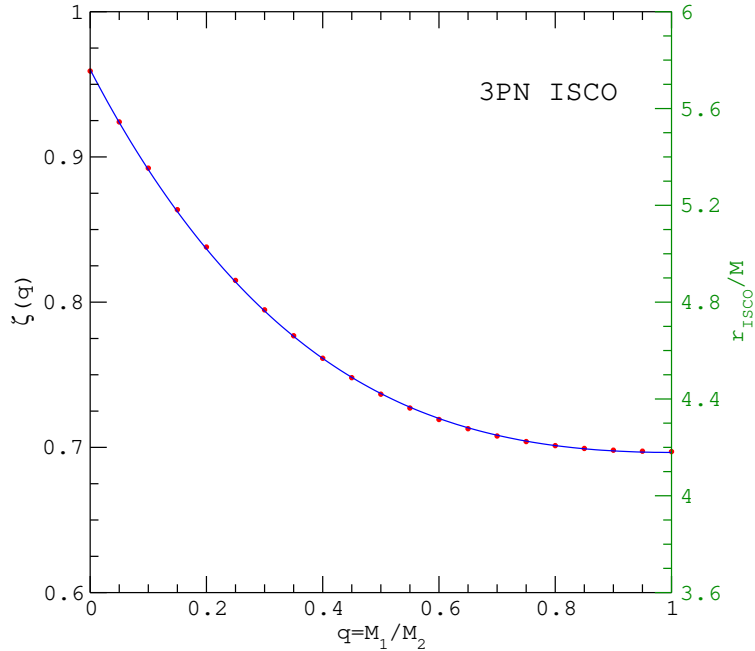}
\caption
[The hybrid corrections $\zeta(q)$ for the Paczy\'nski--Wiita potential.]
{The ISCO radius and the correction function $\zeta(q)$ given
by 3PN results. The correction function $\zeta(q)$ as determined by
solving the cubic equation (\ref{EQ:ZETACUBIC}) are shown as dots
while the fitted function given by equation (\ref{eq:zeta_fit}) is
represented by the solid line.
\label{fig:zeta}}
\end{center}
\end{figure}
The radius of the ISCO $r_{\scriptscriptstyle ISCO}$ to order 3PN, the
correction function $\zeta(q)$, and the fitted correction function
$\zeta_{\scriptscriptstyle FIT}(q)$ are shown in Figure
\ref{fig:zeta}.  We note that $\zeta_{\scriptscriptstyle FIT}(q)$ does
not perfectly reproduce the general relativistic result ({\it i.e.,}
$\zeta(0)=1$) in the limit of small $q$.  This is a consequence of the slow
convergence of the post-Newtonian approach, which is here truncated at
the 3PN order.

\subsection{Roche lobe radii with analytical fits}
\label{SEC:PWROCHE}

Differences between the Paczy\' nski--Wiita potential and the hybrid
potential predictions for Roche lobe radii are minimal. The
Paczy\' nski--Wiita results can be obtained by setting $\zeta(q)=1$ in
the expressions utilized for the hybrid potential. Hence, we
establish expressions for the hybrid case only.

Beginning from the hybrid pseudo-GR potential $\phi_{\sss H}$ in
equation (\ref{eq:hybrid}) that describes the orbital evolution, we
assume that the same potential describes the motion of test particles
in the vicinity of two orbiting massive objects in general relativity.
We establish equipotential surfaces using the same methodology as
\citet{KOPAL1}, \citet{PACZYNSKI1}, and \citet{EGGLETON1}, who studied
the Newtonian Roche geometry.  In the potential $\phi_{\sss H}$, the
motion of either mass $M_i$ with position vector ${\vec r}_i$ in the
co-rotating frame is governed by
\begin{eqnarray}
  M_i\left( \frac{d^2 {\vec r}_i}{dt^2} \right)_{rot}
  & = &  M_i\left( \frac{d^2 {\vec r}_i}{dt^2} \right)_{inertial} -~~ 
   M_i {\vec \omega} \times ({\vec \omega}\times {\vec r}_i)
\end{eqnarray}
where $i=1,\, 2$.  The force on either body in the inertial frame is
given by the gravitational attraction of the potential 
in equation (\ref{eq:hybridpot}). Setting the acceleration in the rotating
frame to zero,
the angular frequency of the binary, using the potential $\phi_H$, is
given by
\begin{eqnarray}
  \omega^2 & = & \frac{M}{a[a-\zeta(q)r_{\sss G}]^2}\, ,
\label{EQ:WHYBRID}
\end{eqnarray}
where $a=|{\vec r}_1-{\vec r}_2|$ is the distance between them.  The
Newtonian result $\omega^2 = M/a^3$ obtains in the limit $r_{\sss
G}/a\rightarrow0$.   In a similar fashion, the position $\vec r$ of a 
third body, with mass $m$ negligible compared to
the two large masses $M_1$ and $M_2$, in the corotating frame is governed
by 
\begin{eqnarray}\label{eq:mpot}
  \left( \frac{d^2 {\vec r}}{dt^2} \right)_{rot}
  & = &  \left( \frac{d^2 {\vec r}}{dt^2} \right)_{inertial}- 
   {\vec \omega} \times ({\vec \omega}\times {\vec r}) - 2  {\vec
  \omega} \times \frac{d{\vec r}}{dt} \, .
\end{eqnarray}
The right hand side includes both the centrifugal force and the
Coriolis force.  Inserting the acceleration using the hybrid potential
in equation (\ref{eq:hybridpot}) in equation (\ref{eq:mpot}) and
integrating, the effective potential in the rotating frame is given by
\begin{eqnarray}\label{eq:rochepotential}
  \phi^{rot}_{\sss H}(x, y) 
&=& -\frac{M}{a}\left[ \frac{x_2}{\sqrt{(x+x_1)^2+y^2}-x_2 z}
  + \frac{x_1}{\sqrt{(x-x_2)^2+y^2}-x_1 z}
  + \frac{1}{2}\frac{x^2+y^2}{(1-\zeta(q)z)^2}\right] \,,
\end{eqnarray}
where we have introduced the dimensionless variables
\begin{eqnarray}
  x=\frac{r_x}{a},\quad && y=\frac{r_y}{a}, \qquad z = 
\frac{r_G}{a} = 2 \frac{M}{a}   \nonumber \\
  x_1=\frac{M_2}{M} = \frac {1}{1+q}\,\quad && 
x_2=\frac{M_1}{M}=qx_1 \,.
\end{eqnarray}
Note that $x_1$ ($x_2$) denotes the position of the mass $M_1$ ($M_2$)
on the negative (positive) dimensionless x-axis.  $z$ is a relativity
parameter.
In writing the above equation, we used the facts that
$\zeta(m/M_1)\rightarrow1$ and $\zeta(m/M_2)\rightarrow1$ for the
negligible mass $m$ of the test particle.  Note that the effects of
the test particle on the total gravitational potential and on the
orbital angular frequency $\omega$ are taken to be negligible.
\begin{figure}[!ht]
\begin{center}
\includegraphics[width=0.975\textwidth]{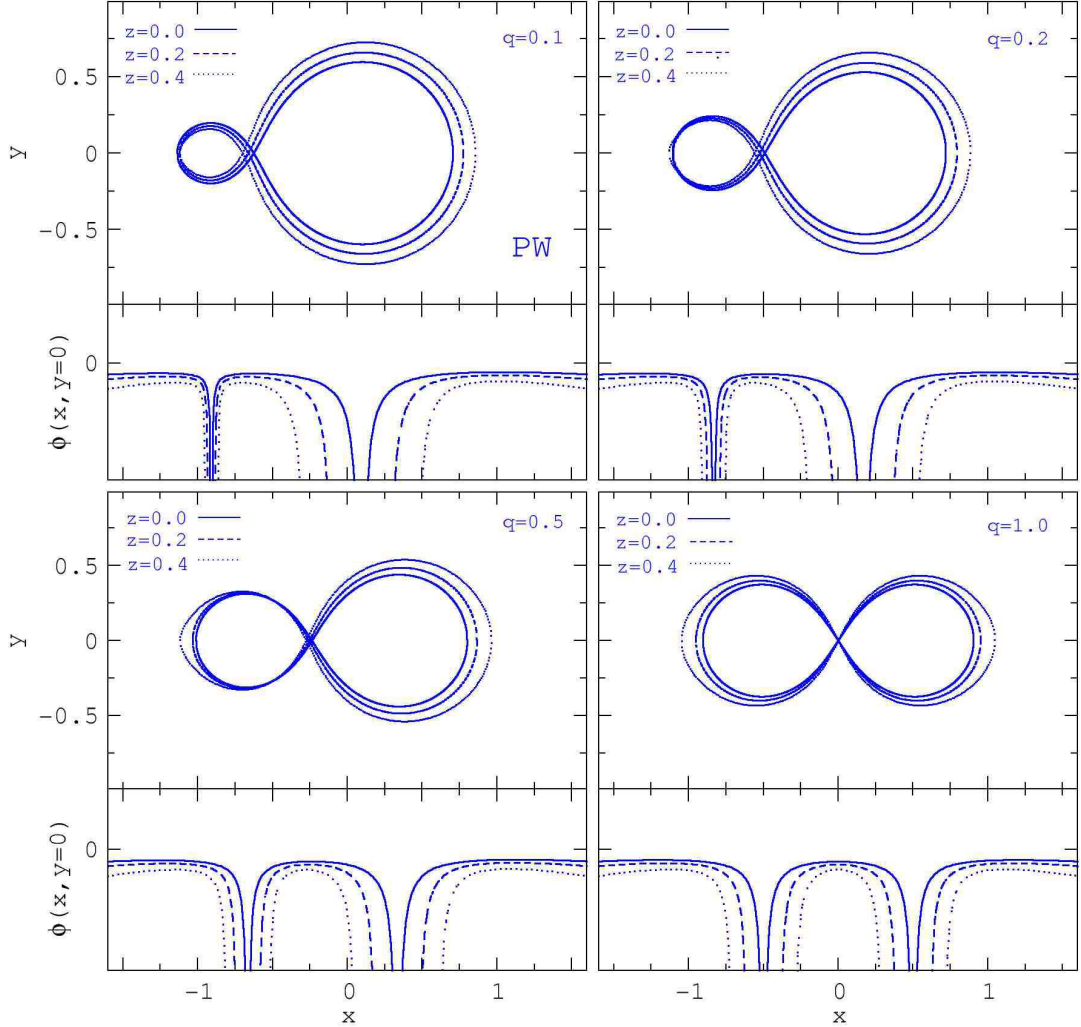}
\caption
[Roche lobes and effective potentials for the Paczy\'nski--Wiita
potential.]
{Dependence of the Paczy\' nski-Wiita potential and its Roche
lobes on $q$ and $z$. \label{fig:rochepotpw}}
\end{center}
\end{figure}
\begin{figure}[!ht]
\begin{center}
\includegraphics[width=0.975\textwidth]{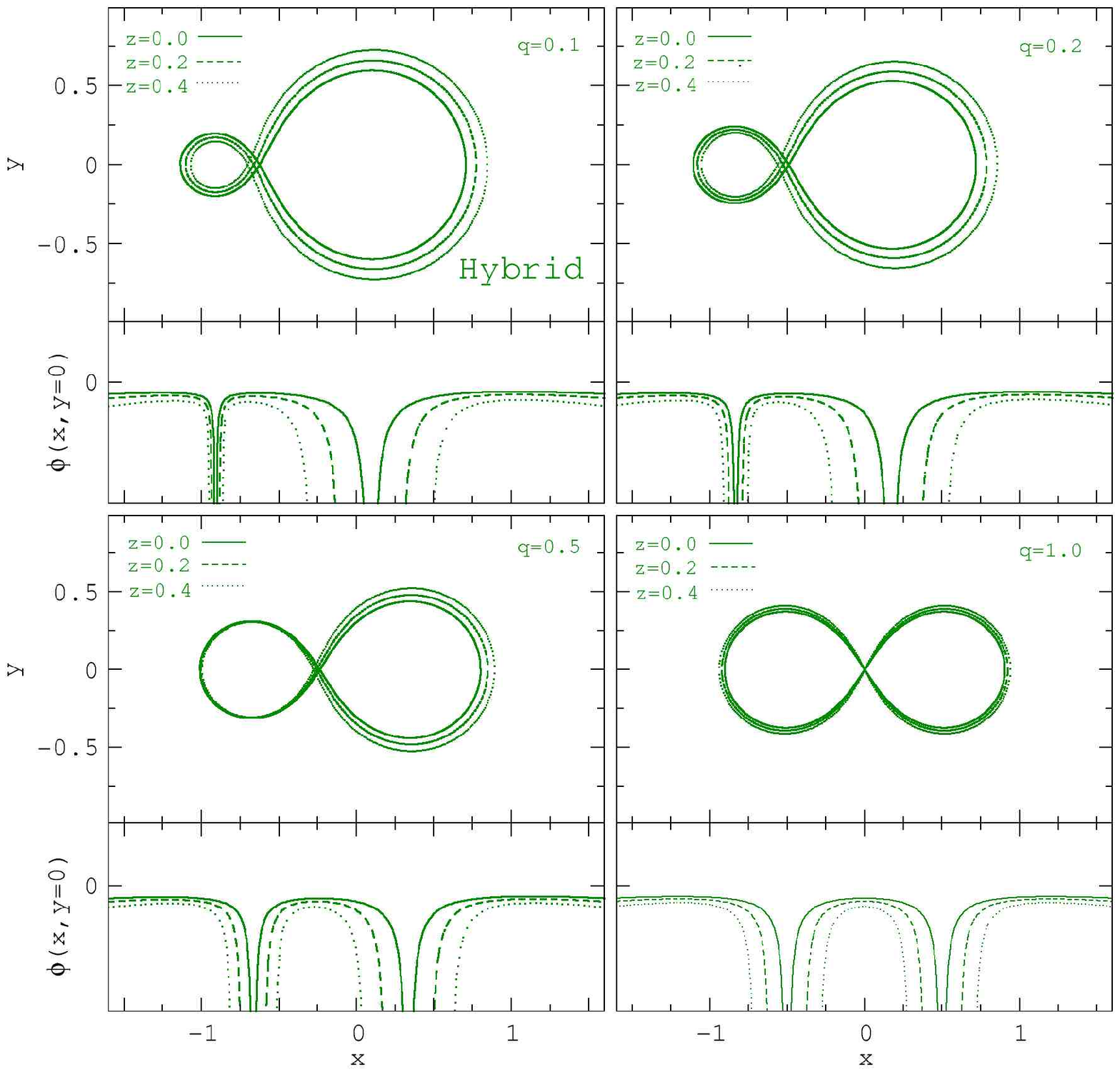}
\caption
[Roche lobes and effective potentials for the hybrid potential.]
{Dependence of the hybrid pseudo-GR potential and its Roche
lobes on $q$ and $z$. \label{fig:rochepothyb}}
\end{center}
\end{figure}

By using equation (\ref{eq:rochepotential}), it is now possible to find
the Roche lobes for the Paczy\' nski--Wiita and for the hybrid cases.
Figures \ref{fig:rochepotpw} and \ref{fig:rochepothyb} display
contours of the effective potentials $\phi^{rot}_{\sss PW}(x, y)$ and
$\phi^{rot}_{\sss H}(x, y)$, and its Roche lobes for four values of
$q$ and three values of $z$.  In contrast to the Newtonian case, for
which the Roche lobe radii depend only on the parameter $q$, the Roche
lobe radii for the two pseudo--potentials depend on the additional
dimensionless parameter $z=r_{\sss G}/a$.

For finite $z$, the hybrid potential in equation (\ref{eq:hybridpot})
causes an effective increase in the strength of gravity compared to
the Newtonian case.  This decreases the value of the potential at the
inner Lagrange point $L_1$ for all values of $q$.  For every $q$, the
inner Lagrangian point moves closer to the lighter mass as $z$
increases.  For a given value of $q$, the volume contained within the
equipotential surface defined by the value of the potential at $L_1$,
i.e., the Roche lobe volume, changes monotonically with $z$.  For
$q<(\beta_C/\alpha_C)^5$, which is 0.63 (0.39) in the case of the
hybrid (Paczy\'nski-Witta) potential, the Roche lobe volume decreases
with increasing relativity parameter $z$.  The opposite behavior is
found for larger values of $q$.

\begin{figure}[!th]
\begin{center}
\includegraphics[width=0.6\textwidth]{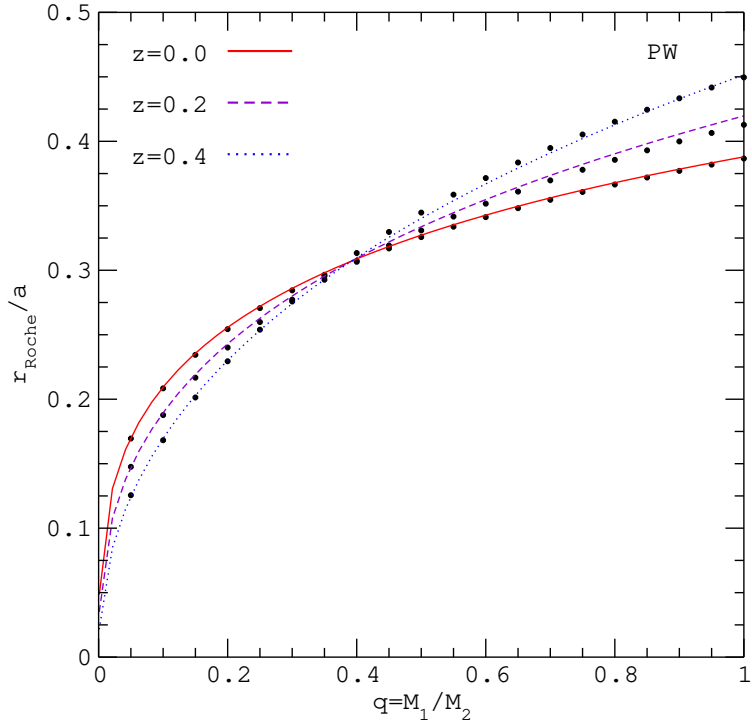}
\caption
[The effective Roche lobe radii for the Paczy\'nski--Wiita potential.]
{The ratio $r_{Roche}/a$ as a function of $q$ and $z$ for the
Paczy\'nski--Wiita potential. The results of numerical
integration are shown by dots whereas the curves represent the
approximation in equation (\ref{eq:approx}). \label{fig:ra}}
\end{center}
\end{figure}

\begin{figure}[!th]
\begin{center}
\includegraphics[width=0.6\textwidth]{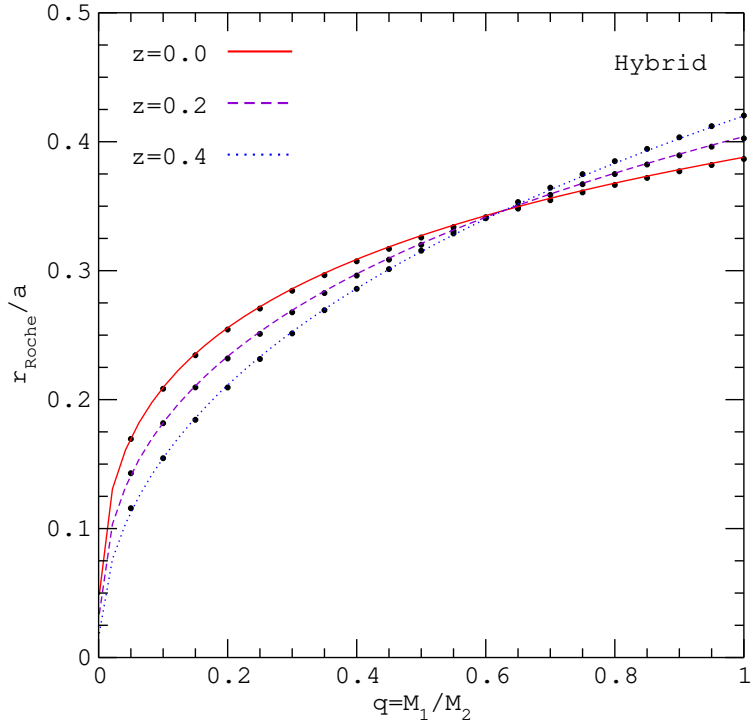}
\caption
[The effective Roche radius for the hybrid potential.]
{The ratio $r_{Roche}/a$ as a function of $q$ and $z$ for the
hybrid potential. The results of numerical integration are represented
by dots whereas the curves represent the approximations in equation
(\ref{eq:approx}). \label{fig:rahyb}}
\end{center}
\end{figure}

After performing numerical integration of the Roche lobes we find that
the Roche lobe radii for both potentials can be adequately described by
using the same functional form as in equation (\ref{eq:reqn}) with
$Q(q)$ given by equation (\ref{eq:roche2pn}); the parameters
$\alpha_C$ and $\beta_C$ for the hybrid and for the PW potentials are
listed in Table \ref{TBL:ROCHEFIT}. For comparison, parameters for the
2PN potential are also given in this table. Numerical determinations of
the Roche lobe radii are compared with these fits in Figures
\ref{fig:ra} and \ref{fig:rahyb}.  The case $z=0$ corresponds to the
Newtonian case studied by \citet{EGGLETON1}.

\begin{table}[th!]
\begin{center}
\begin{tabular}{l c c }
\hline
Potential &$\alpha_C$& $\beta_C$ \\
\hline
\hline
Paczy\' nski--Wiita & 2.398 & 1.988 \\
Hybrid & 2.328 & 2.122 \\
2PN & 1.951 & 1.812 \\
\hline
\end{tabular}
\caption
[Values for the fitting parameters in equation (\ref{eq:c}).]
{Values for the fitting parameters in equation (\ref{eq:c}) for the
potentials that are used in sections \ref{sec:pseudopotentials} and
\ref{sec:evolution} for modeling binary evolution with mass
transfer. The fitting results for the Paczy\' nski--Wiita 
and the hybrid pseudo-GR potentials are depicted in Figures
\ref{fig:ra} and \ref{fig:rahyb}, whereas the second order
post--Newtonian results are taken from \citet{ROCHE2PN}.
\label{TBL:ROCHEFIT}
}
\end{center}
\end{table}

These calculations are restricted to values of $z$ limited by the
condition that the ISCO radius remains smaller than the separation
distance. Quantitatively, this condition amounts to
$z<\zeta(q)^{-1}/3\simeq0.33-0.5$.  Secondly, for large $z$ and small
$q$, the inner Lagrange point $L_1$ becomes the global extremum of the
potential. Consequently, the stars cannot remain in contact and mass
overflow from the lighter star does not flow through $L_1$ to the
other star. In such a situation, the Roche lobe analysis cannot be
used to describe mass transfer. Therefore, we restrict ourselves to
$0\le z \le 0.5$.  We also note that the hybrid potential becomes
unreliable for large $z$ as it lacks the full content of GR.

\subsection{Evolution equations}
\label{sec:NPWHYB}

The energy of a binary system is carried away through its gravity wave
luminosity
\begin{eqnarray}\label{eq:gwluminosity}
  L_{GW} & = & 
  \frac{1}{5}\langle {\reduce \tripledot{I}}_{jk}  {\reduce \tripledot{I}}_{jk} \rangle
  = 
  \frac{32}{5}a^4\mu^2\omega^6\, .
\end{eqnarray}
Simultaneously, its angular momentum diminishes at the rate given by
\begin{eqnarray}\label{eq:gwangmomentum}
  \left(\dot{J}_{GW}\right)_i & = & 
  \frac{2}{5}\epsilon_{ijk} 
  \langle {\reduce \doubledot{I}}_{jm}  {\reduce \tripledot{I}}_{km} \rangle
  = 
  \frac{32}{5}a^4\mu^2\omega^5\, .
\end{eqnarray}

We note that these expressions do not depend on the form of the
gravitational potential involved (i.e. Newtonian, Paczy{\'
n}ski-Wiita, or our hybrid potential). The only assumptions that are
made are weak gravitational field (linearized gravity) and slowly
evolving circular orbits ($a$ and angular frequency $\omega$ do not
change significantly over a single period).

For the hybrid potential in equation (\ref{eq:hybridpot}), the energy
of a binary system in a circular orbit is 
\begin{eqnarray}
  E & = & \frac{a\mu M}{2(a-\zeta(q)r_{\sss G})^2} 
- \frac{\mu M}{a-\zeta(q)r_{\sss G}} \,.
\end{eqnarray}
In the absence of mass transfer, this expression can be used to find the
evolution of the system with time.  As the separation
$a$ is decreases due to the emission of gravitational waves, the
change in energy with time from equation (\ref{eq:gwluminosity}) is
given by 
\begin{eqnarray}
  \dot{E} & = & \frac{\mu M}{2} 
  \frac{a - 3\zeta(q)r_{\sss G}}{(a-\zeta(q)r_{\sss G})^3} \dot{a} 
  = - \frac{32}{5} \mu^2 M^3 \frac{a}{(a-\zeta(q)r_{\sss G})^6}\, .
\end{eqnarray}

The orbital angular momentum of a binary system
with angular frequency $\omega$, separation $a$, and 
reduced mass $\mu$ is $J=a^2\mu\omega$. With the hybrid potential
$\phi_H$, the orbital angular frequency $\omega$ is given by equation
(\ref{EQ:WHYBRID}).  The orbital angular momentum will change in time
because of gravitational radiation reaction.  If mass transfer occurs,
we will assume it conservative so that no further change to the
orbital angular momentum occurs.  Mass transfer, however, will result
in a change in the mass ratio $q$.  In response to radiation reaction
and mass transfer, the change in the angular momentum can be written
as

\begin{eqnarray}
 \dot{J} & = & \frac{d}{dt}\left[\frac{\mu a^{3/2} M^{1/2}}{a-\zeta(q)r_{\sss G}}\right] 
 =
 J\left\{
   \left[\frac{1-q}{1+q}+\frac{r_{\sss G}\, q\, \zeta^\prime(q)}{a-\zeta(q)r_{\sss G}}\right] 
   \frac{\dot{q}}{q}
   + \frac{a - 3\zeta(q)r_{\sss G}}{2(a-\zeta(q)r_{\sss G})}\frac{\dot{a}}{a} \right\}\, ,
\end{eqnarray}
where we have used
\begin{eqnarray}
\dot{A}\equiv \frac{d}{dt}A, \qquad A^\prime \equiv \frac{d}{dq}A\, ,  \qquad {\rm and} \qquad 
  \frac{d\mu}{\mu} = \frac{1-q}{1+q}\,\frac{dq}{q}\, ,
\end{eqnarray}
for $A$ any function of $q$ and/or $t$.
We recognize this as the pseudo-GR version of equation
(\ref{EQ:BASICJ}).  The rate of angular momentum loss is given by
equation (\ref{eq:gwangmomentum}), which we denote as $\dot J_{GW}$.
This can be written as
\begin{eqnarray}\label{eq:aqdiffeq}
  \left[\frac{1-q}{1+q}+\frac{r_{\sss G}\, q\, 
\zeta^\prime(q)}{a-\zeta(q)r_{\sss G}}\right] 
   \frac{\dot{q}}{q}
   + \frac{a - 3\zeta(q)r_{\sss G}}{2(a-\zeta(q)r_{\sss G})}
\frac{\dot{a}}{a}   
  & = &  
  - \frac{\dot{J}_{GW}}{J} = -\frac{32}{5}  
M^2 \frac{\mu(t)}{(a(t)-\zeta(q)r_{\sss G})^4}\,.
\end{eqnarray}
In the absence of mass transfer, $\dot{q}(t)=\dot{\mu}(t)=0$, whence
\begin{eqnarray}\label{eq:diffeqanmt}
  \dot a & = & -\frac{64}{5} ~\frac{a(t)}{a(t) -
  3\zeta(q)r_{\sss G}} ~\frac{M^2 \mu}{(a(t)-\zeta(q)r_{\sss G})^3}\, ,
\end{eqnarray}
shows the orbital decay.

In the event of stable mass transfer, the radius of the lighter star
is pinned to its effective Roche lobe radius, $R_1(t)=r_{\sss
Roche}(t)$.  The time evolutions of $R_1$ and $r_{\sss Roche}$ are
then governed by $q(t)$ and $a(t)$ which, combined with equation
(\ref{eq:roche2pn}) gives the equivalent of equation
(\ref{EQ:BASICROCHE}):
\begin{eqnarray}\label{eq:roche_q}
  \frac{\dot{r}_{\sss Roche}}{r_{\sss Roche}} & = & 
  \left(1 + \frac{a}{C(q, z)} \frac{\partial C(q, z)}{\partial z}
  \frac{\partial z}{\partial a} \right) \frac{\dot{a}}{a}
  + \left(\frac{q}{Q(q)}\frac{\partial Q(q)}{\partial q} +
  \frac{q}{C(q, z)} \frac{\partial C(q, z)}{\partial q} \right)
  \frac{\dot{q}}{q}\, .
\end{eqnarray}
Utilizing the explicit dependences on $q$ and $z$ in equations
(\ref{eq:roche2pn}) and (\ref{eq:c}), the derivatives
needed above are easily evaluated. Explicitly,
\begin{eqnarray}\label{eq:partial}
  \frac{\partial \ln Q(q)}{\partial \ln q} & = &
  \frac{2 \ln(1 + q^{1/3}) - (1 + q^{-1/3})^{-1}}
  {3 [\beta_Q q^{2/3} + \ln(1 + q^{1/3})]} \, , \\
  \frac{\partial C(q, z)}{\partial q} =
  \frac{\alpha_C}{5} \, \frac{z}{q^{4/5}} \, , &\textrm{and}&
  \frac{\partial C(q, z)}{\partial z} =
  \alpha_C\, q^{1/5} - \beta_C\, .
\end{eqnarray}

Equations (\ref{eq:roche_q}) and (\ref{eq:alphadef}) yield an expression 
analogous to equation (\ref{EQ:QDOT}):
\begin{eqnarray}\label{eq:qageneral}
  \frac{\dot{q}}{q} & = &
  \frac{1 -{\displaystyle \frac{\partial\ln C(q, z)}{\partial \ln z}} }
       {\displaystyle \frac{\alpha}{1+q} - \frac{qQ^\prime(q)}{Q(q)}
	 - \frac{\partial\ln C(q, z)}{\partial \ln q}}
       \cdot \frac{\dot{a}}{a}\equiv\Upsilon(a,q)  \frac{\dot{a}}{a}\, ,
\end{eqnarray}
with
\begin{eqnarray}\label{eq:chi2}
  \Upsilon(a,q) & \equiv & 
     \Bigg[
	 \left(1 + z \, (\alpha_C\, q^{1/5} - \beta_C)\right) \nonumber \\
	 && \times \,
	 \left(
	 \frac{\alpha(q, M)}{1+q} - 
	 \frac{2\ln(1+q^{1/3})-(1+q^{-1/3})^{-1}}
	      {3[\beta_Q\, q^{2/3}+\ln(1+q^{1/3})]}
	 \right)
	 -\frac{\alpha_C}{5}\, z\, q^{1/5}
	 {\Bigg]}^{-1}\, .
\end{eqnarray}
Equation (\ref{eq:aqdiffeq}) allows us to separate
the time derivatives of $a$ and $q$ to form the coupled equations 
\begin{eqnarray}\label{eq:diffeqa}
  \frac{d a(t)}{dt} & = &
    -\frac{32}{5}\,\frac{M^3}{a^3}\,\frac{1}{{\left(1-\zeta(q)\, z\right)}^4}\,\frac{q}{(1+q)^2}
    \frac{1}{\displaystyle \left[\frac{1-q}{1+q}
	+ \frac{z\, q\, \zeta^\prime(q)}{1-\zeta(q)\, z} \right]\, \Upsilon(a,q) 
      + \frac{1-3\zeta(q)z}{2(1-\zeta(q)z)}} 
\end{eqnarray}
and
\begin{eqnarray}\label{eq:diffeqq}
  \frac{d q(t)}{dt} & = &
 -\frac{32}{5}\,\frac{M^3}{a^4}\,\frac{1}{{\left(1-\zeta(q)\, z\right)}^4}\,\frac{q^2}{(1+q)^2}
    \frac{\Upsilon(a,q)}{\displaystyle \left[\frac{1-q}{1+q}
	+ \frac{z\, q\, \zeta^\prime(q)}{1-\zeta(q)\, z} \right]\, \Upsilon(a,q) 
      + \frac{1-3\zeta(q)z}{2(1-\zeta(q)z)}}\,. 
\end{eqnarray}
Note that in the absence of mass transfer ($\Upsilon(a,q)=0$),
equation (\ref{eq:diffeqa}) reduces to equation (\ref{eq:diffeqanmt}),
and equation (\ref{eq:diffeqq}) yields $\dot{q}(t)=\dot{\mu}(t)=0$.
We note that the relativistic modifications complicate the evolution
equations, but they do not qualitatively alter them.

\section{Conditions for stable mass transfer}

During the early stages of binary evolution, mass transfer is
negligible and the orbital evolution is independent of the
EOS. However, as the two objects approach each other, the Roche lobe
radius shrinks until it becomes equal to the stellar radius $R_1$.  At
this time, the lighter companion overfills its Roche lobe and mass
transfer becomes possible.  However, it is not guaranteed that mass
transfer will be stable.  As shown in \S \ref{sec:General}, the
condition for stable mass transfer is equation (\ref{EQ:STABLE}).
If this condition is satisfied, mass
transfer proceeds smoothly.  If this condition is violated, mass
transfer either terminates or grows uncontrollably since the star's radius
increases faster than the Roche lobe radius.
The restrictions imposed by equation (\ref{EQ:STABLE}) can be shown
graphically for each choice of the gravitational potential and EOS.
However, it is useful to examine some limiting situations with a
somewhat simpler analytic expression for $Q(q)$, derived by 
\citet{KOPAL1}, so that a qualitative interpretation can be made. To
begin with, we neglect effects of relativity so that $z=0$.   In this
case, the $a$ dependence in $\Upsilon$ can be ignored.  Using the
simple expression 
\begin{eqnarray}
\label{eq:kopal}
Q(q)
&=&
0.46224\,
{
\left(
\frac{q}{1+q}
\right)
}^{1/3}\, ,
\end{eqnarray}
which \citet{KOPAL1} found to be a good approximation for $q<0.8$, we
obtain the approximate relation
\begin{eqnarray}
\Upsilon(q) 
&\simeq&
\frac{1+q}{\alpha-1/3}\, . 
\label{EQ:UPSILONN}
\end{eqnarray}
For realistic neutron star or strange quark matter equations of state,
$\alpha\le 1/3$, so that $\Upsilon$ is negative as inferred in \S
\ref{sec:General}. In the case of a more detailed Roche lobe fit (like
equation (\ref{eq:q})), the limit to $\alpha$ can be somewhat smaller,
which sometimes precludes very low mass strange quark stars from
continuously transfering its mass in a stable fashion onto its
companion.

\begin{figure}[!thb]
\begin{center}
\includegraphics[width=0.55\textwidth]{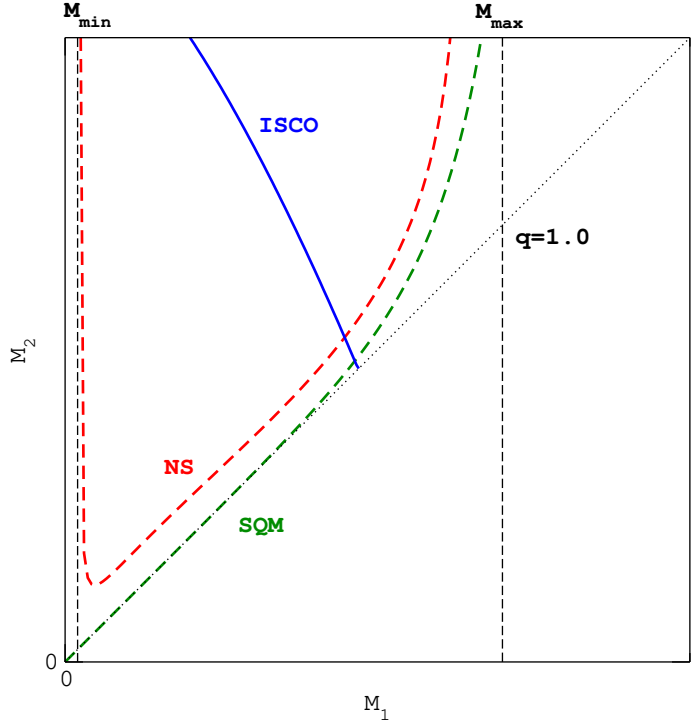}
\caption
[The stable mass transfer and ISCO constraints.]
{The limiting
condition, that is, the lower mass limit to $M_2$, 
for stable mass transfer is indicated by a dashed curve
for the neutron star (NS) and strange quark matter star (SQM) cases.
The solid curve shows the upper mass boundary for mass transfer
beginning outside the ISCO. Other lines show the condition $q=1.0$ and the
maximum and minimum neutron star masses, $M_{max}$ and $M_{min}$,
respectively.
\label{FIG:STABLEISCOSKETCH}}
\end{center}
\end{figure}

In the Newtonian case, equation (\ref{EQ:STABLE}) now becomes \begin{eqnarray}
\Upsilon(q)
<
-\frac{1}{2}\, \frac{1+q}{1-q}\, .
\label{EQ:QCONDN}
\end{eqnarray}
Utilizing equations (\ref{EQ:UPSILONN}) and (\ref{EQ:QCONDN}) turns the
condition in equation (\ref{EQ:STABLE}) into
\begin{eqnarray}
q
<
\frac{\alpha(M_1)}{2} + \frac{5}{6}\, ,
\qquad{\textrm{or}}\qquad
\alpha(M_1)
>
2\, q - \frac{5}{3}\, .
\label{EQ:SIMPLESTABLE}
\end{eqnarray}
By selecting a fixed value for $q$, we can use this simple formula to
find the range of masses for which a given equation of state allows
for stable mass transfer to occur. In Figure \ref{fig:rmalpha}, we
designate with horizontal lines three values for $q$ that are of
special interest. For $q=0$, corresponding to a heavy binary
partner or a small $M_1$, we obtain a line with $\alpha_{\sss
0}=-5/3$. The equal mass $q=1$ case is denoted with $\alpha_{\sss
1}=1/3$ in this figure, and the case $q=0.5$ (often used in 
numerical evolutions in later sections) with $\alpha_{\sss 0.5}=-2/3$. 
We note that for mass transfer to be stable, the actual
$\alpha(M_1)$ has to be greater than these values. In general, 
this forbids stable mass transfer for the $q=1$ case even for the strange
quark matter EOS,
except in the $M_1\rightarrow 0$ limit. Generally, stable mass
transfer will be possible for normal EOS's when $q$ is farther from
unity than for strange quark mass stars. The limit for stable mass transfer
imposed by equation (\ref{EQ:STABLE}) is shown schematically by the dashed
curves for normal and SQM stars in Figure \ref{FIG:STABLEISCOSKETCH}.
We note that since the EOS
parameter $\alpha$ becomes very negative for normal stars with low
mass (see Figure \ref{fig:rmalpha}), stable mass transfer eventually
becomes impossible.  This occurs for $M_1$ slightly larger than 
$M_{min}\simeq0.09$ M$_\odot$, the minimum stable neutron star mass.  
This lower mass restriction, however, does not exist in the case of 
SQM stars.

The assumption of quasi-circular evolution in the pseudo-GR potential
cannot be applied beyond the ISCO, the point of gravitational
instability after which direct plunge onto the heavier companion is
inevitable.  Although mass transfer might ultimately ensue during the
plunge, our analysis is not trustworthy should mass transfer
begin after the ISCO is reached.  This establishes the limiting condition
\begin{eqnarray}\label{eq:upper}
  R_1\left(M_1\right)  =  r_{\sss ISCO}\, Q(q)\, C(q, {\tilde z})\, ,
\end{eqnarray}
with $\tilde{z}=2M/r_{\sss ISCO}$, which can be phrased as a
constraint in terms of $M$ and $q$, or, alternatively, in terms of
$M_1$ and $M_2$.  We utilize this condition for all potentials,
despite the fact that the Newtonian potential does not predict an ISCO
and the PW potential is deeper than the 2PN potential.
The solid curve in Figure \ref{FIG:STABLEISCOSKETCH} shows, schematically, 
the limit dictated by equation (\ref{eq:upper}): stable mass transfer 
occurs outside the ISCO for ($M_1,M_2$) values below this curve. 
For sufficiently large $M_2$, or for
$M_1\simeq M_{max}$, both
normal and self-bound stars are unable to transfer mass in a stable
fashion within the ISCO limitation.  From Figure
\ref{FIG:STABLEISCOSKETCH}, we infer that there exists a window in the
$M_1$--$M_2$ plane in which a merging binary system could evolve with stable
and continuous mass transfer.

\begin{figure}[!htb]
\begin{center}
\includegraphics[width=1.0\textwidth]{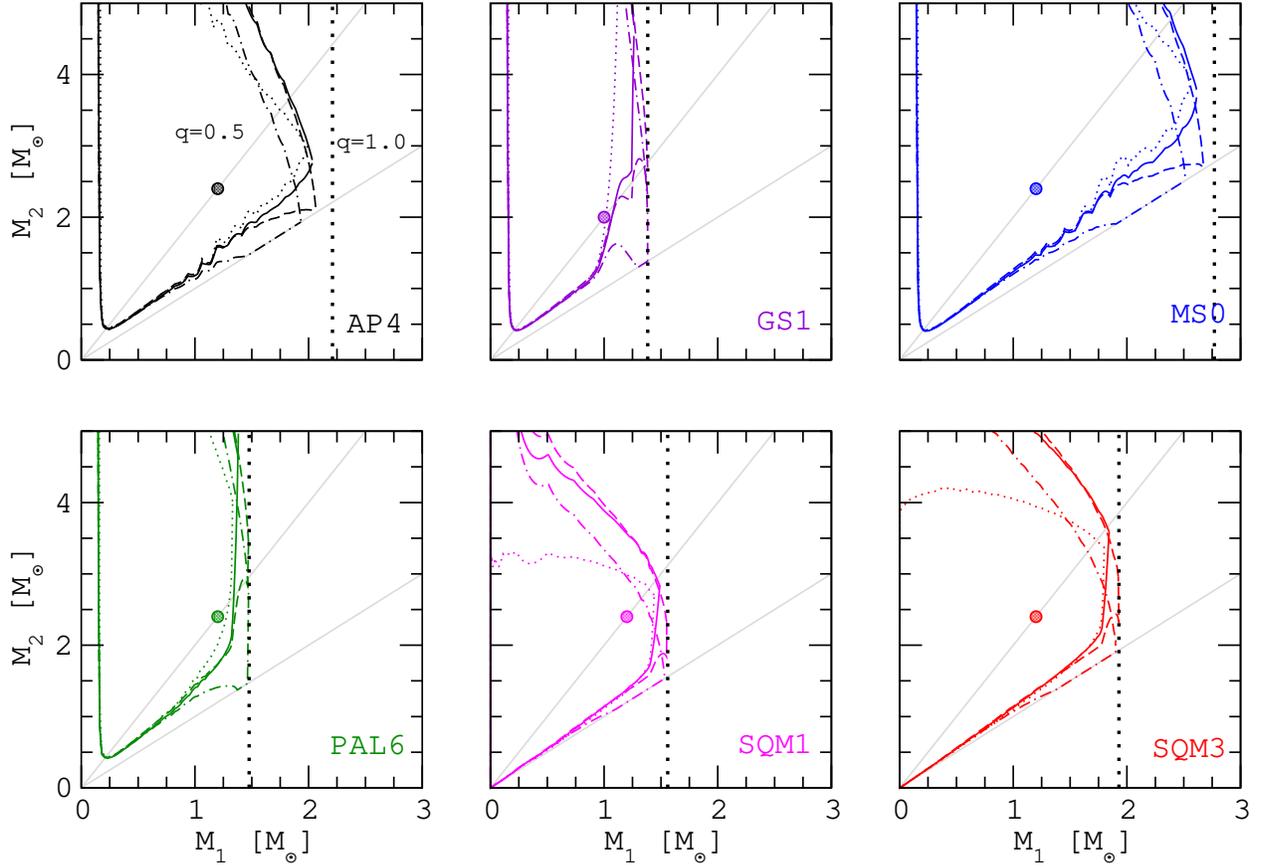}
\caption
[The stable mass transfer and ISCO constraints for real EOSs.]
{Limiting masses for stable mass transfer combining equations
(\ref{EQ:STABLE}) and (\ref{eq:upper}) for several equations of state
and for the 2PN, hybrid pseudo-GR, Paczy\'nski-Witta and Newtonican
potentials, denoted by solid, dashed, dot-dashed and dotted curves,
respectively.  Filled circles indicate the masses employed in
simulations discussed in \S \ref{sec:evolution}.  Two additional lines
that correspond to $q=1.0$ (equal mass case) and $q=0.5$ (used in our
evolutions) are shown.  }
\label{fig:stable}
\end{center}
\end{figure}

We now combine equations (\ref{EQ:STABLE}) and (\ref{eq:upper}) with
the results for all four potentials and for all six equations of state
that we have used (see Table 1 for notation and references). The
ensuing results are shown in Figure \ref{fig:stable}. The qualitative
picture of our simple analysis summarized in Figure
\ref{FIG:STABLEISCOSKETCH} is not modified, only the quantitative
details change.  As in Figure \ref{FIG:STABLEISCOSKETCH}, we find
regions in the $M_1$--$M_2$ plane for all four EOS's for which stable
mass transfer is possible.  In the next section, we discuss the
evolution of merging binaries that undergo stable mass transfer.


\section{Results of model evolutions}
\label{sec:evolution}

The numerical integration of the set of coupled differential equations
in equations (\ref{EQ:DIFFA}) and (\ref{EQ:DIFFQ}) were performed
using a fourth order Runge-Kutta algorithm.  The drastically different
$\alpha(M)$ functions for normal and strange quark matter stars
produce rather different behaviors after stable mass transfer
begins. As a result, gravity wave emissions from these two types of
mergers also differ. To underscore this, we will calculate the scalar
gravitational polarization amplitude
\begin{eqnarray}\label{eq:h}
h_+(t) &=& \frac{4}{r}\, \frac{M^2}{a}\, \frac{q}{(1+q)^2}\, 
\cos 2\omega(t-r)\, ,  
\end{eqnarray}
where $r$ is the distance from the binary system to the observer. The observed frequency of the emitted gravitational waves will be
$2\times 10^5~\omega/\pi$ Hz.

\begin{figure}[ht!]
\begin{center}
\includegraphics[width=0.9\textwidth, angle=0]{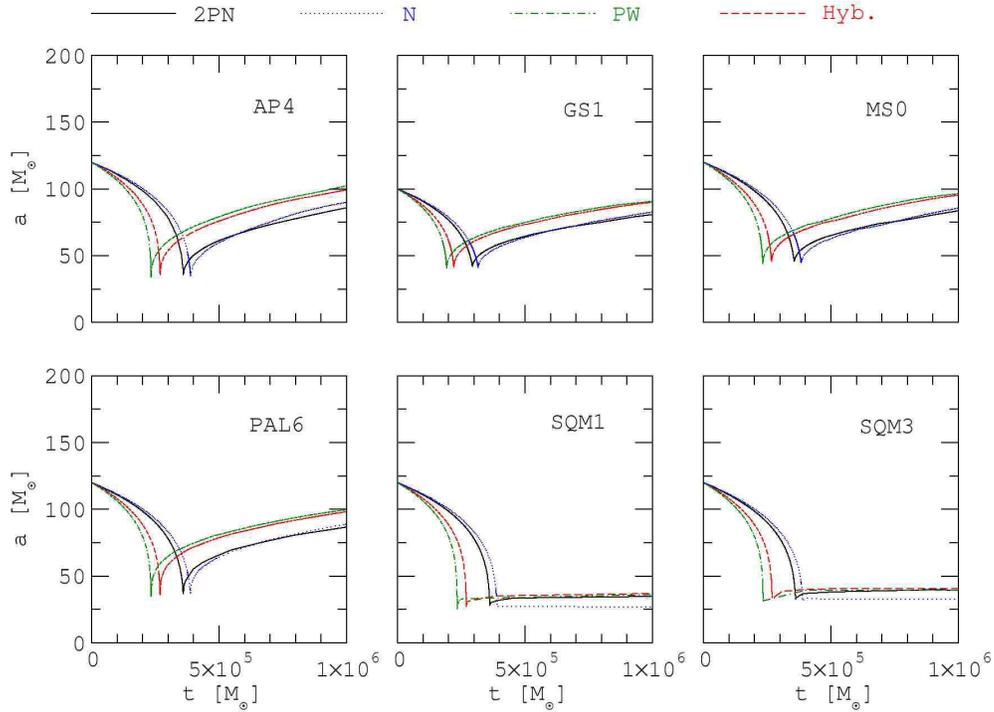}
\caption
[Evolution of the radial separation $a(t)$.]
{Evolution of the semimajor axis $a$ ($1$ $M_\odot=1.475$ km, $1$
$M_\odot=4.92\times 10^{-6}$ s) for the six EOS's of Table 1.  Line
designations are as in Figure \ref{fig:stable}.
\label{FIG:EVOLUTIONA}}
\end{center}
\end{figure}

\begin{figure}[ht!]
\begin{center}
\includegraphics[width=0.9\textwidth, angle=0]{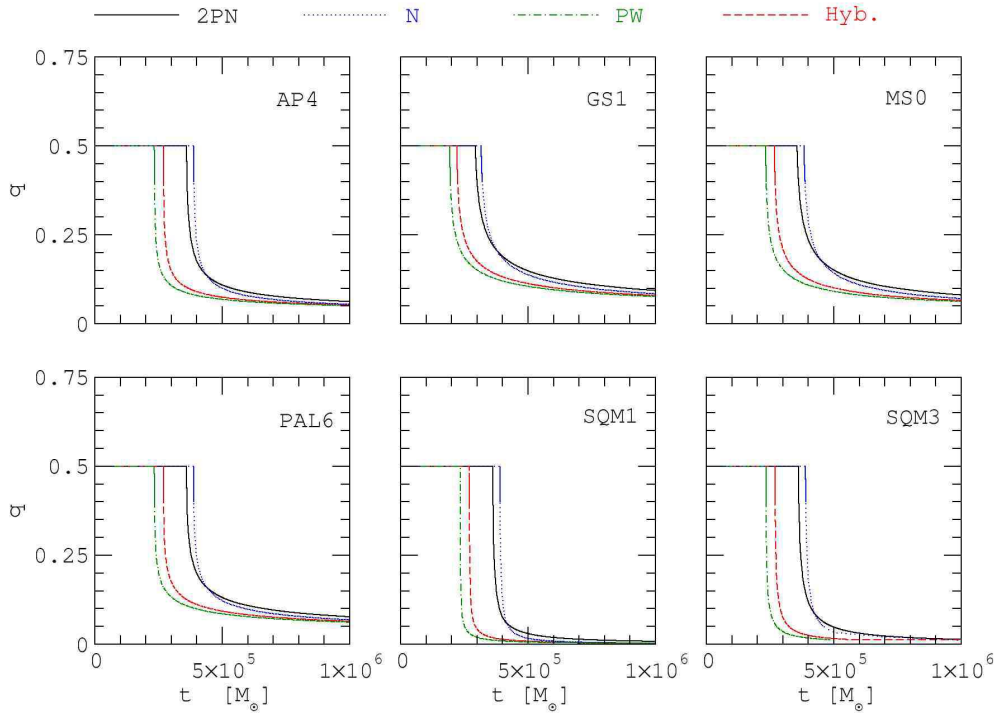}
\caption
[Evolution of the mass ratio $q(t)$.]
{The same as Figure \ref{FIG:EVOLUTIONA}, but for the mass ratio $q$. 
\label{FIG:EVOLUTIONQ}}
\end{center}
\end{figure}

\begin{figure}[ht!]
\begin{center}
\includegraphics[width=0.9\textwidth, angle=0]{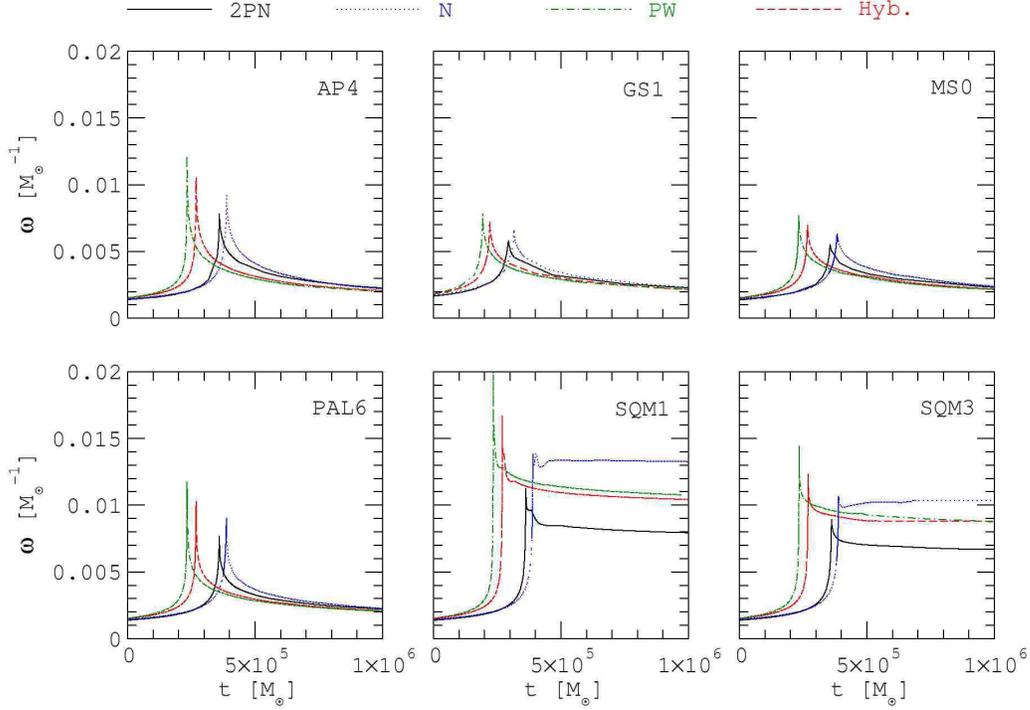}
\caption
[Evolution of the angular frequency $\omega(t)$.]
{The same as Figure \ref{FIG:EVOLUTIONA}, but for the angular frequancy $\omega$.  Note that the observed gravitational wave frequency is $\omega/\pi$.
\label{FIG:EVOLUTIONW}}
\end{center}
\end{figure}

\begin{figure}[ht!]
\begin{center}
\includegraphics[width=0.9\textwidth, height=0.675\textwidth, angle=0]{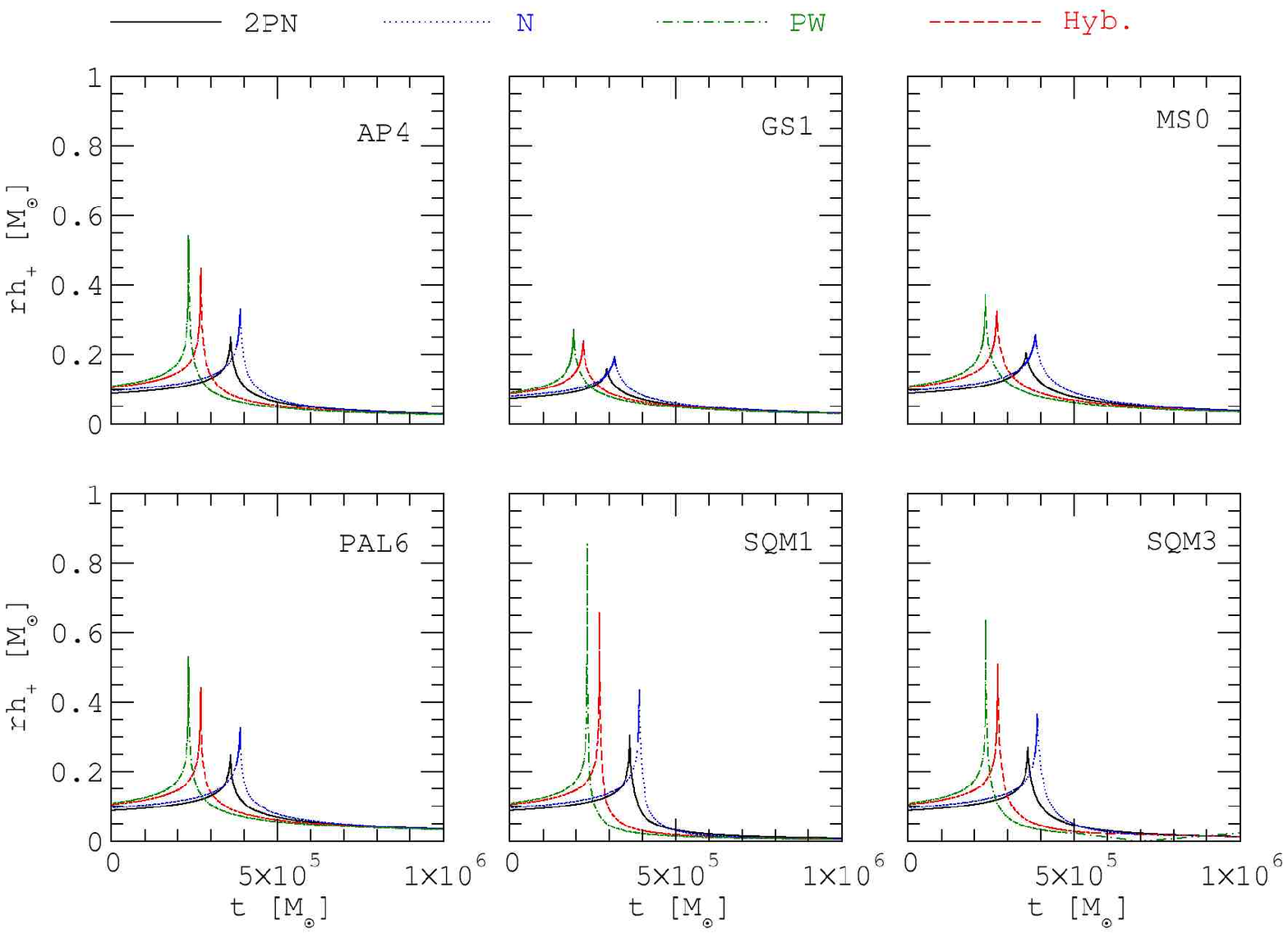}
\caption
[Evolution of the amplitude of the gravitational stress $h_{+}(t)$.]
{The same as Figure \ref{FIG:EVOLUTIONA}, but for the amplitude of the
  gravitational stess $|h_+|$ multiplied by the distance $r$.
\label{FIG:EVOLUTIONH}}
\end{center}
\end{figure}

Figures \ref{FIG:EVOLUTIONA}--\ref{FIG:EVOLUTIONH} compare the mergers
of normal and self-bound stars with a BH as described by equations
(\ref{eq:diffeqa}) and (\ref{eq:diffeqq}). For these examples, the
initial neutron star mass was taken to be 1.2 $M_\odot$ in all cases
except that of EOS GS1, for which the initial neutron star mass was
taken to be 1.0 M$_\odot$.  In all cases, the initial mass ratio was
$q=0.5$, which guarantees that an epoch of stable mass transfer
results.  We are primarily interested in detailing the effects of
stable mass transfer, so we do not consider cases in which tidal
disruption occurs within the ISCO.  For each EOS, evolutions were
generated for the various gravitational potentials considered in this
paper: Newtonian, second order post-Newtonian, and the two pseudo-GR
potentials.  Inspiral is characterized by increases in the orbital
frequency $\omega$ and scalar gravitational polarization amplitude
$h_+$, a decrease in orbital separation $a$, and a fixed $q$.  Stable
mass transfer ensues at the ``kinks'' visible in the evolution of
these quantities.  Within each class of EOS, i.e., normal and
self-bound stars, variations in the EOS only qualitatively alter the
results.  During stable mass transfer, the decrease in orbital
separation and rise in frequency and waveform amplitudes are reversed.
The major effect of incorporating general relativistic corrections to
the potential is to speed up the evolution relative to the Newtonian
case.  Stable mass transfer thus begins earlier in these cases.  GR
corrections also result in a somewhat larger value for the orbital
separation following the onset of mass transfer.

Large differences are apparent between the evolutions of the normal
and self-bound cases.

\begin{enumerate}

\item The orbital separation $a(t)$ increases after mass transfer
begins in the normal neutron star case, but quickly saturates in the
self-bound case;

\item Reflecting the behavior of $a$, following the onset of stable
  mass transfer, the orbital angular frequency $\omega(t)$
  continuously decreases in the normal case, but quickly achieves a
  relatively constant value in the self-bound case.  A comparison of
  the evolution of $\omega$ for different equations of state is
  displayed in Figure \ref{fig:allomegas} for the 2PN case, in which
  the zeros of time have been adjusted so that the onset of mass
  transfer is simultaneous.  Also included in this figure are results
  for a polytropic equation of state $ P=K\rho^\Gamma$ with $K =
  0.0445 c^2/\rho_n$, $\rho_n = 2.3\times 10^{14}~{\rm g~cm^{-3}}$,
  and $\Gamma = 2$, which yields a radius of $\simeq 20.4$ km for the maximum
  mass star of 1.2M$_\odot$.  For the polytropic equation of state,
  $\alpha\rightarrow0$ as $M\rightarrow0$.  Relatively large
  differences exist among the SQM star cases compared to the normal
  neutron star cases.

\begin{figure}[ht!]
\begin{center}
\includegraphics[width=0.6\textwidth, angle=0]{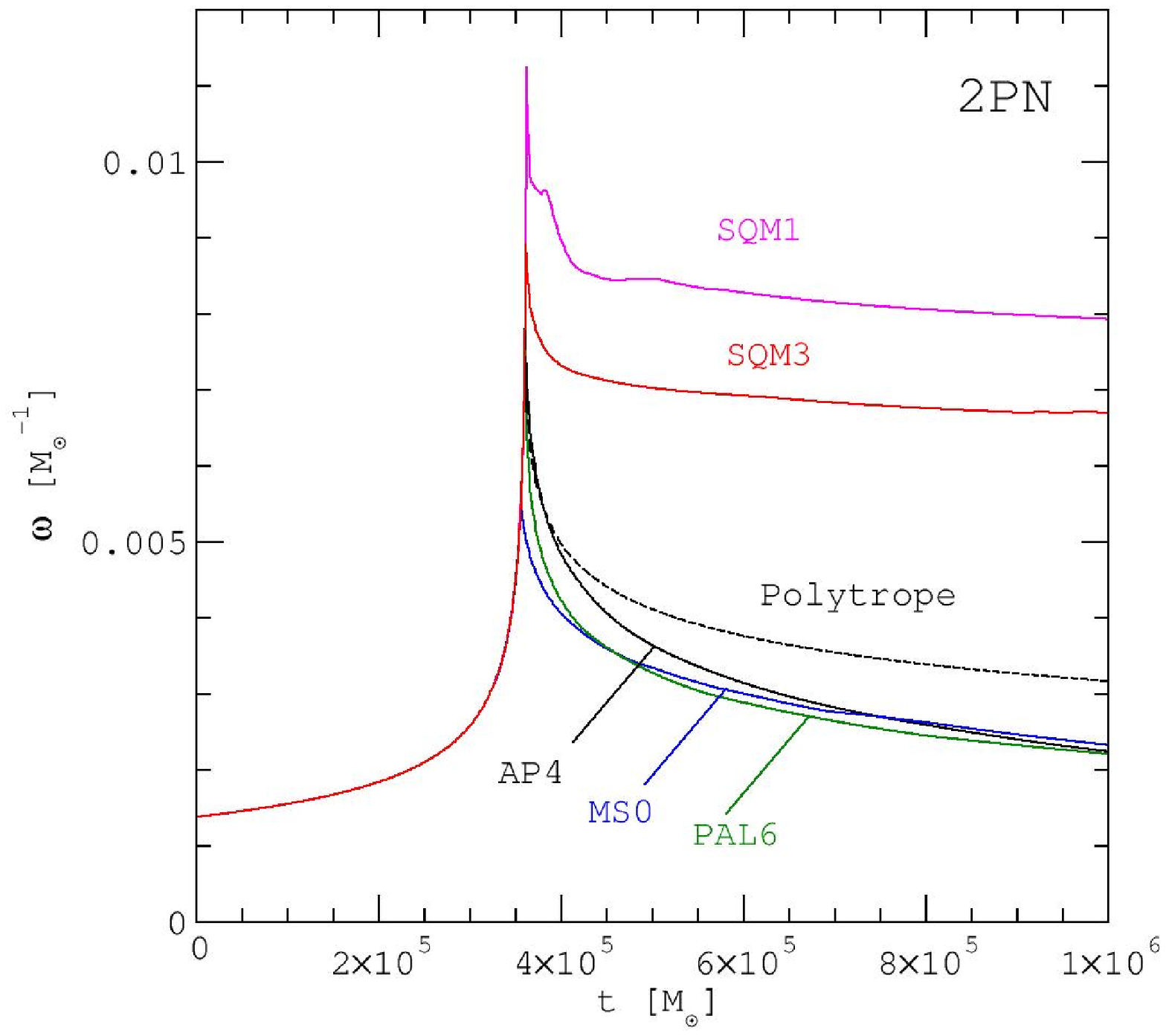}
\caption
[Evolution of $\omega$.]
{The evolution of $\omega$ for different equations of state for the
  Newtonian case, with the zeros of time adjusted so that the onset of
  mass transfer is simultaneous.  
\label{fig:allomegas}}
\end{center}
\end{figure}

\item In the normal neutron star case, the mass $M_1$ approaches about
  0.16 M$_\odot$.  Eventually, stable mass transfer terminates when
  $\alpha$ becomes too negative.  Further evolution cannot be followed
  realistically with our model, but what apparently occurs is that the
  stellar radius now increases faster than the Roche lobe radius,
  indicating a catastrophic evaporation of mass from the neutron star.
  Since this occurs so closely to the neutron star minimum mass, about
  0.09 M$_\odot$ (this is independent of the supra-nuclear EOS), this
  evaporation is quickly followed by the decompression of the
  remaining neutron star \citep{COLPI93}.  In contrast, the mass of
  the self-bound star dwindles to extremely small values.  In this
  case, the minimum mass limit is that of a strange quark nugget,
  determined in part by surface and Coulomb effects.

\item The envelope of the gravitational waveform amplitude $|h_+(t)|$
follows the behavior of $q$ and $a$.  In the normal neutron
star case, the amplitude decreases steadily until stable mass transfer
terminates.  Following an initially rapid decrease, the amplitude
decreases very slowly with time and continues indefinitely
in the quark star case.

\end{enumerate}


\begin{figure}[ht!]
\begin{center}
\includegraphics[width=0.6\textwidth, angle=0]{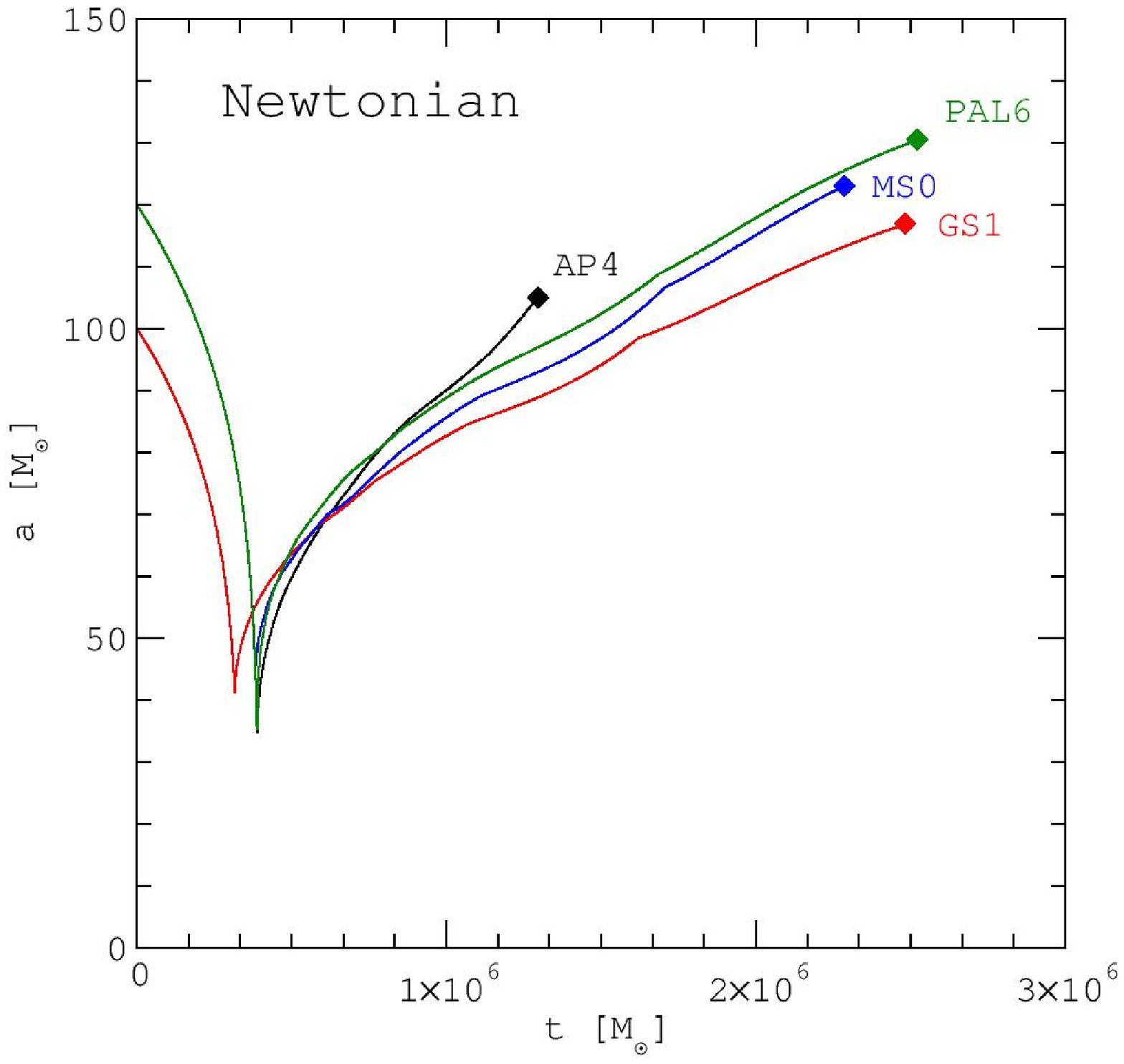}
\caption
[Full evolution of the separation $a(t)$.]
{The evolution of the semimajor axis $a(t)$ for the Newtonian case,
followed up to the end of stable mass transfer (indicated by filled
diamonds), for the neutron star EOS's AP4, PAL6, MS0 and GS1.
\label{FIG:FULLA}}
\end{center}
\end{figure}

The evolution during the stable mass transfer phase can be
qualitatively understood.  The large differences observed between the
strange quark star and the normal star cases arise as a result of the
differing behaviors of the $\alpha(M_1)$ function in the limit of
small $M_1$.  For the SQM case, $\alpha(M_1\rightarrow0)=1/3$, whereas
in the normal star case, it becomes increasingly negative.  From
equation (\ref{eq:chi2}), $\Upsilon$ quickly increases and tends to
infinity for small $q$ in the SQM case, so that $\dot a\rightarrow0$
and $a$ quickly stabilizes at a fixed value not much different than
its value when stable mass transfer begins.  However, in the normal
star case, $\Upsilon$ steadily decreases, so that the denominators in
equations (\ref{eq:diffeqa}) and (\ref{eq:diffeqq}) tend to zero as
$q$ decreases.  Stable mass transfer terminates when the denominator
vanishes, and formally $\dot q$ and $\dot a$ tend to infinity.
However, 
the temporal evolution is speeded up so
that both $a$ and $q$ reach particular final values in a finite time
as shown in Figure (\ref{FIG:FULLA}).  Mass continues to be lost from
the neutron star, but now in a catastrophic, unstable, fashion because
the stellar radius is expanding faster than the Roche lobe.  When the
star's mass reaches $M_{min}$, the star essentially explodes as it
becomes gravitationally unbound.  The decompressing matter can then be
expected to achieve relatively large velocities, possibly exceeding
escape velocity from the system.

We can analytically examine the evolution in the normal star case
by making a few simplifying approximations.  First, we note that
$\alpha$ diverges as the minimum stable neutron star mass, $M_{min}$,
is approached.  A useful approximation, valid for small $M_1$, is
\begin{equation}
\label{eq:alphaapprox}
\alpha\approx-{\frac{\alpha_0}{M_1-M_{min}}}\,,
\end{equation}
where $\alpha_0\simeq0.089$ and $M_{min}\approx0.09$ M$_\odot$.  Using
the Kopal approximation for $Q(q)$, equation (\ref{eq:kopal}), stable
mass transfer terminates when
\begin{equation}
\label{eq:alphaf}
\alpha_f=2q_f-{\frac{5}{3}}=-{\frac{\alpha_0(1+q_f)}{q_f(M-M_{min})-M_{min}}}\,.
\end{equation}
This suggests, for our standard case $M=3.6$ M$_\odot$, that
$\alpha_f\simeq-1.59$ and $q_f\simeq0.04$, independently of the EOS.
These values are close to those obtained in the numerical
integrations.  Furthermore, if equation (\ref{eq:alphaapprox}) is
also utilized, one can analytically integrate equation
(\ref{eq:qageneral}) to find
\begin{equation}
\label{eq:aaf}
{\frac{a}{a_f}}\simeq\left({\frac{q_f}{q}}{\frac{1+q}{1+q_f}}\right)^
{1/3}
\left({\frac{1+q_f}{1+q}}{\frac{\alpha}{\alpha_f}}\right)^{\alpha_0/M_{min}}\,.
\end{equation}
Although equation (\ref{eq:alphaapprox}) is not valid for large values
of $M_1$, using that relation results in an estimate for our standard
case that $a/a_f\simeq0.21$, where $a$ refers to the orbital
separation when mass transfer begins.  For comparison, Figure
(\ref{FIG:FULLA}) indicates that $a/a_f$ is actually in the range of
0.25--0.33.  There is no simple way to estimate the duration of
stable mass transfer, but figure (\ref{FIG:FULLA}) indicates it is of
order $1-2\times10^6$ M$_\odot$ or about 5--10 seconds.

Similarly, we can analytically express the evolution of an SQM star
during stable mass transfer using the Kopal approximation, leading to
\begin{eqnarray}
\label{eq:sqmapprox}
\dot a&=&-{\frac{32}{5}}\left(\frac{M}{a}\right)^3\frac{q}{(1+q)^2}
\left(\frac{\alpha-1/3}{\alpha/2+5/6-q}\right)\,,\cr
\dot q&=&-\frac{32}{5}\frac{M^3}{a^4}\frac{q^2}{1+q}\left(\frac{1}{\alpha/2+5/6-q}\right)\,.
\end{eqnarray}
Since $\alpha=1/3$ for moderate to small SQM star masses, we may
assume $a$ is approximately constant and equal to the value
at which mass transfer begins.  The equation for $\dot q$ is then
integrated to yield
\begin{equation}
\label{eq:deltatm}
\frac{\Delta t}{M}\simeq\frac{5}{32}\left(\frac{a}{M}\right)^4\left(\frac{1}{q_f}+q_f-\frac{1}{q}-q\right)\,,
\end{equation}
where $q_f$ is the value of $q$ achieved for a time $\Delta t$ after
the stable mass transfer begins.  Since the gravitational wave
amplitude $h_+$ varies as $q/(1+q)^2$, as seen from equation
(\ref{eq:h}), we expect this amplitude to decrease with time as
$t^{-1}$, for an indefinite period of time.

However, how small can an SQM star become before the bulk EOS for SQM
is inadequate?  The bulk EOS is valid for an infinitely large
system. As the strange quark star becomes small, Coulomb and surface
effects become increasingly important. As shown by \citet{FARHI84},
the bulk EOS is accurate only for values of the total baryon number
$A=(N_u + N_d + N_s)/3>10^7$, which in our case translates to a value
for $q_f\approx2\times10^{-51}$.  Obviously, $q>>q_f$, so from
equation (\ref{eq:deltatm}) one finds $\Delta t\approx10^{41}$ yr.
Therefore the final disruption or merger of the strange quark matter
star could never be observed.


One way to gauge the effects of different gravitational potentials is
to compare the orbital evolution during inspiral.  In order of
relative gravity strengths, or attractiveness, we note the potential
ordering Newtonian, 2 PN, hybrid and PW from Figures
\ref{FIG:EVOLUTIONA}--\ref{FIG:EVOLUTIONH}.  The time elapsed during
inspiral can be found by integrating equation
(\ref{eq:diffeqanmt}), which results in
\begin{eqnarray}\label{eq:aa}
  (a_{\sss fin}^4 - a_{\sss ini}^4) - 8 \zeta(q)r_{\sss G} (a_{\sss
  fin}^3 - a_{\sss ini}^3) + 24 (\zeta(q)r_{\sss G})^2 (a_{\sss fin}^2
  - a_{\sss ini}^2) && \nonumber \\ - 40 (\zeta(q) r_{\sss G})^3
  (a_{\sss fin} - a_{\sss ini}) + 12 (\zeta(q) r_{\sss G})^4
  \ln(a_{\sss fin}/a_{\sss ini}) & = & - \frac{256}{5}\mu M^2 (t_{\sss
  fin} - t_{\sss ini})\, ,
\end{eqnarray}
where the initial radial separation and time are $a_{\sss ini}$ and
$t_{\sss ini}$, respectively, and the final separation and time
($a_{\sss fin}$ and $t_{\sss fin}$, respectively) are chosen at the
moment when mass transfer begins.  The difference in the timescales
between two potentials, for example, Newtonian and PW, can be
established by differencing equation (\ref{eq:aa}) for the two cases.
However, although $a_{\sss ini}$ and $t_{\sss ini}$ can be taken as
the same for the two cases, the two values of $a_{\sss fin}$ will be
slightly different.  Nevertheless, to lowest order this difference can
be neglected and the difference $\Delta t=t_{\sss fin, N}-t_{\sss fin,
PW}$ in evolution times for the two cases can be expressed as an
expansion in the small quantity $\zeta(q)r_G$:
\begin{equation}\label{eq:deltat}
\frac{\Delta t}{M}\simeq\frac{5}{16} \frac{\zeta(q)r_g(1+q)^2}{q}
\left(\frac{a_{\sss fin}^3-a_{\sss ini}^3} {M^3}\right)+\mathcal{O}[(\zeta(q)r_G)^2]\,.
\end{equation}
For our simulations, $M=3.6$ M$_{\odot}$, $t_{\sss ini}=0$ and $a_{\sss
ini}=120$ M$_\odot$, whence $\Delta t\simeq1.5\times10^5$ M$_\odot$.
This agrees well with our numerical result.

\section{Discussion}
\label{sec:Discussion}

As discussed in \citet{CUTLER94}, careful analysis of the
gravitational waveform during inspiral yields values for not only the
chirp mass $M_{chirp}=(M_1M_2)^{3/5}/M^{1/5}$, but for also the
reduced mass $M_1M_2/M$, so that both $M_1$ and $M_2$ can be found.
Thus, observation of stable mass transfer effects in the gravitational
wave signal will allow several details about neutron star structure to
be discerned, including several features of the underlying EOS.  For
example,

\begin{enumerate}

\item The condition $r_{Roche}=R_1$ at the onset of mass transfer will
allow a direct estimation of $R_1$ as $a$ at that point is determined
by $M$ and $\omega$, $a=M^{1/3}/\omega^{2/3}$.  For example, in the
Newtonian case, using the Kopal relation for $Q$, equation
(\ref{eq:kopal}), this relation predicts that $\omega R_1=0.46^{3/2}
M_1^{1/2}.$   Thus a point ($M_1,R_1$) on the mass-radius diagram can be
estimated.  The effects of variations due to neutron star radii are
less subtle for stable mass transfer than those which influence the
inspiralling gravitational waveform \citep{FABER02}.

\item Similarly, the condition $r_{Roche,f}=R_{1f}$ at the end of stable mass
transfer yields a relation between $M_{1f}$ and $R_{1f}$, where the
subscript $f$ refers to the final values indicating the end
of stable mass transfer.

\item The ratio of the gravitational wave amplitudes $h_+$ at its peak
and at the end of stable mass transfer $h_{+f}$, together with the
ratio $\omega/\omega_f$, then can yield a value for the ratio
$R_1/R_{1f}$.  Since $R_1$ has already been found, it is then feasible
to estimate both $R_{1f}$ and $M_{1f}$ separately, so a second
mass-radius pair can be found.  For example, if the simple Kopal formula
is used for the Roche lobe geometry, one finds
\begin{eqnarray}
\frac{R_1}{R_{1f}}
&=&
\left(\frac{h_+}{h_{+f}}\right)^{1/3}
\left(\frac{\omega_f}{\omega}\right)^{8/9}
\left(\frac{1+q}{1+q_f}\right)^{1/3}\,.
\end{eqnarray}

\item Moreover, the end of stable mass transfer establishes a value
( $\alpha_f$) for $\alpha$ there.  If we again use the simple Kopal
relation for the Roche lobe geometry, equation (\ref{eq:kopal}), we obtain
$\alpha_f\simeq-5/3+2q_f$.  For low-mass neutron stars, $M_1<0.4$
M$_\odot$, we have observed that equation (\ref{eq:alphaapprox})
is valid.  This equation is relatively independent of the high-density EOS
because such low-mass stars are primarily composed of matter below the
nuclear saturation density.  If the value for $\alpha_f$ is consistent
with $q_f$, it means that mass transfer was indeed conservative.  If
they are not consistent, it is still possible, in principle, to then
deduce a value for $M_a$ which can be used to revise the estimates for
$M_{1f}$ and $R_{1f}$ accordingly.

\item The time elapsed between the onset of stable mass transfer and
its termination depends in a complicated way on $q, M, R_1$ and
$\alpha(M)$.  Nevertheless, measurement of this time, assuming that
$q$ and $M$ are already measured, and that $R_1$ can be estimated as
indicated above, implies that the function $\alpha$ can be
constrained.

\item Most importantly, the sharp contrast between the evolutions
during stable mass transfer of a normal neutron star and a strange
quark star should make these cases distinguishable even if other
information concerning the values of $q, M$ and $R_1$ is weak.  This
result is independent of the form of the gravitational potential or
the nuclear or quark matter EOS employed. During the phase in which
mass transfer is present a normal star outspirals in order to preserve
the total angular momentum of the system until the minimum mass limit
is reached. In contrast, a strange quark matter star hovers at
approximately constant separation distance during mass transfer for a
very long time.

\item In addition, for the case of strange quark matter stars, the
differences in the height of the frequency peak and the plateau in the
frequency values at later times are related to the differences in
radii of the stars at these two epochs.  It is, therefore, indirectly
an indicator of the maximum mass of the star: the closer is the star's
mass before mass transfer to the maximum mass, the greater is the
difference between these frequency values, because the radius change
will be larger. As shown in Figure \ref{fig:allomegas}, the SQM1 case
has a larger drop in the frequency than the SQM3 case because 1.2
M$_\odot$ is closer to the maximum mass for the SQM1 case.
Information about the maximum mass could therefore be revealed by
these merging systems.  Together with radius information, the value of
the maximum mass remains the most important unknown that could reveal
the true equation of state at high densities.
\end{enumerate}

Given that the peak frequency observed in the gravitational wave
signal has such a prominent role in deterining the stellar radius in
the event of stable mass transfer, we summarize our results for the
Newtonian case in Table \ref{tbl:WEOS}.  We also indicate the
estimated peak frequency using the approximation of equation
(\ref{eq:kopal}) for the Newtonian case.  The inverse connection
between the peak frequency and the radius is apparent.

\begin{deluxetable}{l l l l l l l}
\tablecaption{The value of the angular frequency $\omega$ at its
peak.}
\tablewidth{0pt}
\tablehead{
EOS& 
\colhead{AP4} & 
\colhead{GS1} &
\colhead{MS0} & 
\colhead{PAL6} & 
\colhead{SQM1} &
\colhead{SQM3}
}
\startdata
$\omega_{\sss \textrm{peak, 2PN}}$ [$10^{-3}$ M$_\odot^{-1}$] & 
$9.20$ & 
$6.54$ & 
$6.28$ & 
$9.02$ & 
$13.83$ & 
$10.65$ \\
$\omega_{\sss \textrm{Eq. (\ref{eq:kopal})}}$ [$10^{-3}$ M$_\odot^{-1}$] & 
$9.25$ & 
$6.56$ & 
$6.32$ & 
$9.13$ & 
$13.79$ & 
$10.78$ \\
$R_1$ [km] & 
11.4 & 
13.5 & 
14.7 & 
11.5 & 
8.7 & 
10.3 
\enddata 
\tablecomments{Comparison of the angular frequencies at the beginning
of mass transfer. In the first line, we show results for the angular
frequencies $\omega$ from our calculations in Figure
\ref{FIG:EVOLUTIONW}. In the second line, we show $\omega$ as given by
the Newtonian approximation utilizing equation (\ref{eq:kopal}). All
cases were computed for $M_1=1.2$ M$_\odot$ and $M_2=2.4$ M$_\odot$,
except GS1 for which $M_1=1.0$ M$_\odot$ and $M_2=2.0$
M$_\odot$. The corresponding radii $R_1$ at those masses are also shown.
\label{tbl:WEOS}}
\end{deluxetable}

\begin{deluxetable}{ccccc}
\tablecaption{The mass transfer-ISCO restriction for
\citet{KLUZNIAK1}. }
\tablewidth{0pt}
\tablehead{ 
\colhead{$q$} & 
\colhead{$M_1/M_\odot$} &
\colhead{$M/M_\odot$} & 
\colhead{$R_1$ [km]} & 
\colhead{$R_{\sss Roche, ISCO}$ [km]} 
}
\startdata 
0.5 & 1.5 & 4.5 & 9 & 8.43 \\ 
0.3 & 1.5 & 6.5 & 9 & 10.17 \\ 
0.5 & 2.0 & 6.0 & 12 & 11.25 \\ 
0.3 & 2.0 & 8.67 & 12 & 13.57 \\ 
0.5 & 2.5 & 7.5 & 15 & 14.05 \\ 
0.3 & 2.5 & 10.83 & 15 & 16.96 \\ 
\enddata
\label{tbl:lee}
\end{deluxetable}

\subsubsection*{Relation with other works}

We can compare our results to hydrodynamical simulations, both
Newtonian and pseudo-GR, that have been performed.  \citet{KLUZNIAK1}
reported a series of hydrodynamical simulations 
of the mergers of a strange quark matter star
with a more massive companion using the PW
pseudo-GR potential.  They found that the strange quark star
plunged for any choice of masses in the binary system. However, the
six cases they studied would, by our analysis, either fall into the
situation $R_1<r_{Roche,ISCO}$, for which mass transfer would begin
only after the SQM star is plunging, or are borderline cases. We can
employ equation (\ref{eq:upper}) to find restrictions analogous to the
ones in Figure \ref{fig:stable} for the cases studied in
\citet{KLUZNIAK1}. In Table \ref{tbl:lee}, we list the computed Roche
radius at the ISCO for the six cases presented in \citet{KLUZNIAK1}
and compare it to the results given there.

From Table \ref{tbl:lee}, we note that the three $q=0.3$ cases fall
into the class of configurations that lead to a plunge.  For the
remaining three cases, we would have predicted stable mass transfer
and a relatively slower subsequent evolution.  Nevertheless, all three
of these cases are close to the borderline of the plunge regime and,
although \citet{KLUZNIAK1} attribute this effect to properties of the
strange quark star, it could well be that the plunge they observed is
primarily a property of the PW pseudo-GR potential, which is somewhat
stronger (deeper) than the 2PN potential we employed.  The ISCO is
closer in for the 2PN potential compared to the PW potential, as
graphically illustrated in Figure \ref{fig:zeta}.

 As noted by \citet{FABER02}, compactness of the star might be
recorded in the emerging gravitational radiation during inspiral. For
the example of a polytropic EOS, the spectrum $dE/df$ of gravitational
waves depends on the ratio of the star's mass and its radius
$M_1/R_1$.  However, in our analysis, the height of the frequency peak
just before mass transfer begins contains important information about
the radius of the star $R_1$ that might be more accurate than that
extracted from the inspiral waveform. Hence, conditional upon having
information about $M_1$ and $M_2$, the frequency of the emerging
gravitational signal could be utilized to measure the star's radius.


Several assumptions have been utilized in our simulations, including
the conservation of the total mass and angular momentum (except for
the angular momentum radiated in gravitational radiation) of the
stars, and quasi-circular orbits Conservation of mass and angular
momentum of the stars could be modified by the formation of an
accretion disc or by the ejection of mass from the system.  In the
absence of a violent tidal disruption, ejection of mass to infinity
seems unlikely, but an accretion disc could form a reservoir that
could modify the orbital evolution during stable mass transfer and the
resulting gravity wave signal.  Also, the assumption of circular
orbits is not reliable near the ISCO, with moderate errors in the
evaluation of the orbital phase (see \citet{MILLER1}).  More accurate
numerical work will be required to establish detailed waveforms for
comparison to experiment, especially if stable mass transfer ensues
near the ISCO.

Another assumption is that the ISCO marks a sharp
cutoff outside of which stable mass transfer is assumed to be
possible.  \citet{MILLER2} argues that
angular momentum loss to gravitational radiation is significant during
the time interval necessary for mass transfer to ensue, so that the
plunge of the stars actually starts well outside the ISCO.  Using the
Peter's formula for angular momentum loss due to gravitational
radiation, and assuming the Newtonian formula for orbital angular
momentum, one can estimate $N(a_i)$, the number of orbits remaining
until the ISCO is reached, as a function of the initial orbital
separation $a_i$:
\begin{eqnarray}
N(a_i) &=& \int_{a_i}^{a_{ISCO}}\frac{1}{ P(a)\dot J_{GW}(a)}
\frac{dJ(a)}{da}da\cr &=& \frac{(1+q)^2}{64\pi q}\left(\frac{a_{ISCO}}{M}\right)^{5/2}
\left[\left(\frac{a_i}{a_{ISCO}}\right)^{5/2}-1\right]
\simeq0.92\left[\left(\frac{a_i}{a_{ISCO}}\right)^{5/2}-1\right]\,.
\label{na}
\end{eqnarray}
In the last line above,  
we used $M_1=1.2$ M$_\odot$ and
$M_2=2.4$ M$_\odot$ and $a_{ISCO}/M\simeq4.4$ as appropriate up to 2PN
order.  From equation (\ref{na}), we find that for $a_i/a_{ISCO}=1.1$ only
1/4 orbit remains until $a=a_{ISCO}$.  If $a_i/a_{ISCO}=1.5$, we find
5/4 orbits remain.  The relevant timescales are determined by the
orbital period, which is approximately $P=0.7 (a/a_{ISCO})^{3/2}$ ms
for our conditions.  Clearly, mass transfer would have to occur
relatively quickly for our scenario to be realized.  Newtonian
SPH hydrodynamical simulations, however, have indeed found
that stable mass transfer can occur (see, {\it e.g.,} \citet{Rosswog04}).

Results using the Newtonian potential are suspect because the
values of the angular frequency and separation near the onset of
stable mass transfer imply orbital velocities in the range 0.5 to
0.6$c$. (See \citet{THORNE1} for more elaboration on this problem.)
Our results indicate, however, that the use of either of the two
pseudo-GR potentials probably overestimates the strengthening of
gravity due to GR. The results for the 2-PN case, in fact, are
intermediate between the Newtonian and pseudo-GR cases and relatively
close to the Newtonian results.

In this context, it is interesting to compare recent calculations of
black hole-neutron star mergers performed by \citet{Rosswog04} and
\citet{Rosswog05}.  In the former paper, Newtonian simulations are
carried out, whereas in the latter paper the Newtonian potential is
replaced by the Paczy\' nski-Wiita potential.  Epsisodic (stable) mass
transfer is found in the Newtonian case whereas the use of the PW
potential results in tidal disruption of the neutron star
inside the ISCO (but accompanied by the ejection of a small-massed
tidal tail).  Our calculations indicate that the Paczy\' nski-Wiita
potential is too strong compared to the 2PN potential, which gives
results closer to the Newtonian case.  Also, the Paczy\' nski-Wiita
simulations of \citet{Rosswog05} only considered black hole masses
greater than 14 M$_\odot$, in which case we would not have expected stable
mass transfer to take place in any event.  

The argument of \citet{MILLER2} concerning the lack of time for mass
transfer to stabilize the orbit of the merging binary is independent
of whether the geometry considered is Newtonian or relativistic.
Since the Newtonian simulations of \citet{Rosswog04} and others ({\it
cf.}, \citet{Lee01} and references therein) showed that stable mass transfer
was nevertheless possible, it would be very interesting to see if
stable mass transfer also occurs in
hydrodynamic simulations that use a potential
correct to 2PN together with smaller black hole masses.

\section{Conclusions}
\label{sec:conclusion}

We have extended Newtonian models of binary orbital evolution and
Roche lobe geometry to second post-Newtonian order to examine the
final stages of compact binary mergers in which the lighter star is a
neutron star or a strange quark matter star.  In our simulations,
binary mergers can be categorized into two distinct classes, depending
upon whether or not stable mass transfer occurs.  We find that
similar, non-negligible, regions in mass space ($M_1-M_2$) allow for
the possibility of stable mass transfer for both normal neutron stars
and strange quark matter stars.  Mass transfer is not expected to
occur if (1) the binary mass ratio $q=M_1/M_2$ is too close to unity,
or (2) the total mass of the system is too large ($M\gtrsim10$
M$_\odot$).  The limit on $q$ for normal stars is $q\gtrsim0.75$, but for
strange quark matter stars it is $q\gtrsim 0.9$.

Binary mergers in which stable mass transfer is able to occur behave
dramatically different than those in which plunge occurs.
Qualitatively, normal stars and SQM stars follow a similar evolution:
stable mass transfer initiates a period of reverse, or outwards,
spiralling, characterized by diminishing frequencies and gravitational
wave amplitude, that has a duration of 5--10 s in the case of normal
stars and lasts essentially forever in the case of SQM stars.  In
contrast, the gravitational wave emission of a plunge is characterized
by a single burst of high-frequency radiation.  These differences
should be 
distinguishable from gravitational radiation
observations.

In the case of stable mass transfer, large differences in the gravity
wave signal following the onset of mass transfer between the cases of
normal neutron stars and strange quark matter stars are expected.
This signature may be unique in its ability to distinguish between
these stellar models from astrophysical observations.  Whereas normal
neutron star evolutions will be characterized by rapidly diminishing
frequencies, strange quark matter star evolutions will have a
radiation frequency that reaches an asymptotic value.  In addition,
stable mass transfer from a normal neutron star will have a finite
duration of order 5--10 s, concluding when the star expands quickly
and overfills its Roche lobe when its mass decreases to about 0.16
M$_\odot$, which will be quickly (of order a few ms) followed by the
violent decompression of the star when it reaches its minimum mass
(about 0.09 M$_\odot$).  In contrast, mass transfer should continue
essentially forever in the case of an SQM star, with a gravitational
wave amplitude that decreases with time as $t^{-1}$.

Several observational constraints become possible
from mergers in which stable mass transfer occurs.  Most important
is the ability to distinguish between the star being normal or
composed of strange quark matter.  In either case, the possibility
exists to extract the radius-mass relation.  Not only will a distinct
$M-R$ point be potentially measureable from a single system,
corresponding to the onset of stable mass transfer, but in the case
that the star is normal, an additional point is possible corresponding
to the termination of stable mass transfer.  The overall signal may
also allow estimation of the function $\alpha(M)=d\ln r/d\ln M$ for
intermediate masses.  Moreover, observations of many different events
will allow sampling of a corresponding number of $M-R$ points.

Research support of the U.S. Department of Energy under grant number
DOE/DE-FG02-87ER-40317 for all authors and grant number
DOE/DE-FG02-93ER-40756 for Madappa Prakash is gratefully acknowledged.

%
%

\bibliographystyle{apj}
\bibliography{references}

\end{document}